\documentclass{aa}     
\usepackage{graphicx}
 
\def\cm2{\,{\rm cm^{-2}}}
\def\kms{\,{\rm {km\,s^{-1}}}}
\def\kkms{\,{\rm {K\,km\,s^{-1}}}}
\def\Msun{\,{\rm M_\odot}}

\def\h2{\,{\rm H_{2}}}
\def\13co{\,{\rm ^{13}CO}}
\def\co{\,{\rm ^{12}CO}}

\def\aua{{\rm A\&A} }
\def\auas{{\rm A\&AS} }
\def\apj{{\rm ApJ} }
\def\aj{{\rm AJ} }
\def\apjs{{\rm ApJS} }
\def\apjl{{\rm ApJL} }

\def\mnras{{\rm MNRAS} }
\def\pasj{{\rm PASJ} }

\begin{document} 
 
\title{Results of the ESO-SEST Key Programme on CO in the Magellanic Clouds}
 
\subtitle{X. CO emission from star formation regions in LMC and SMC} 
 
\author{F.P. Israel
	\inst{1},
        L.E.B. Johansson
	\inst{2},
	M. Rubio,
	\inst{3}
 	G. Garay
	\inst{3},
        Th. de Graauw 
        \inst{4},
        R.S. Booth
	\inst{2}
        F. Boulanger
	\inst{5, 6},
	M.L. Kutner
	\inst{7},
	J. Lequeux
	\inst{8},
	\and 
	L.-A. Nyman
	\inst{2, 9}
} 

\offprints{F.P. Israel} 
 
\institute{Sterrewacht Leiden, P.O. Box 9513, 2300 RA Leiden, the Netherlands 
\and Onsala Space Observatory, S-439-92 Onsala, Sweden
\and Departamento de Astronomia, Universidad de Chile, Casilla 36-D, 
     Santiago, Chile 
\and Laboratorium voor Ruimteonderzoek, SRON, Postbus 800, 9700 AV Groningen,
     the Netherlands
\and Radioastronomie, Ecole Normale Superieure, 24 rue Lhomond, F-75231, Paris 
     CEDEX 05, France
\and Institut d'Astrophysique Spatiale, Bat. 120, Universit\'e de Paris-XI, 
     F-91045 Orsay CEDEX, France
\and Astronomy Department, University of Texas at Austin, USA
\and LERMA, Observatoire de Paris, 61 Av. de l'Observatoire, F-75014 Paris,
     France
\and European Southern Observatory, Casilla 19001, Santiago 19, Chile
}

\date{
Received ????; accepted ????}
 
\abstract{ 
We present $J$=1--0 and $J$=2--1 $\co$ maps of several star-forming 
regions in both the Large and the Small Magellanic Cloud, and briefly
discuss their structure. Many of the detected molecular clouds are 
relatively isolated and quite small with dimensions of typically 20 pc.
Some larger complexes have been detected, but in all cases the
extent of the molecular clouds sampled by CO emission is significantly 
less than the extent of the ionized gas of the star-formation region.
Very little diffuse extended CO emission was seen; diffuse CO in between
or surrounding the detected discrete clouds is either very weak or absent.
The majority of all LMC lines of sight detected in $\13co$ has an
isotopic emission ratio $I(\co)/I(\13co)$ of about 10, i.e. twice
higher than found in Galactic star-forming complexes. At the lowest
$\co$ intensities, the spread of isotopic emission ratios rapidly 
increases, low ratios representing relatively dense and cold molecular 
gas and high ratios marking CO photo-dissociation at cloud edges.
} 

\authorrunning{F.P. Israel et al.}
\titlerunning{CO from Magellanic Cloud HII regions}

\maketitle

\section{Introduction} 

\begin{figure}
\unitlength1cm
\vspace{0.12cm}
\begin{minipage}[t]{8.9cm}
\vspace{0.12cm}
\resizebox{8.9cm}{!}{\rotatebox{270}{\includegraphics*{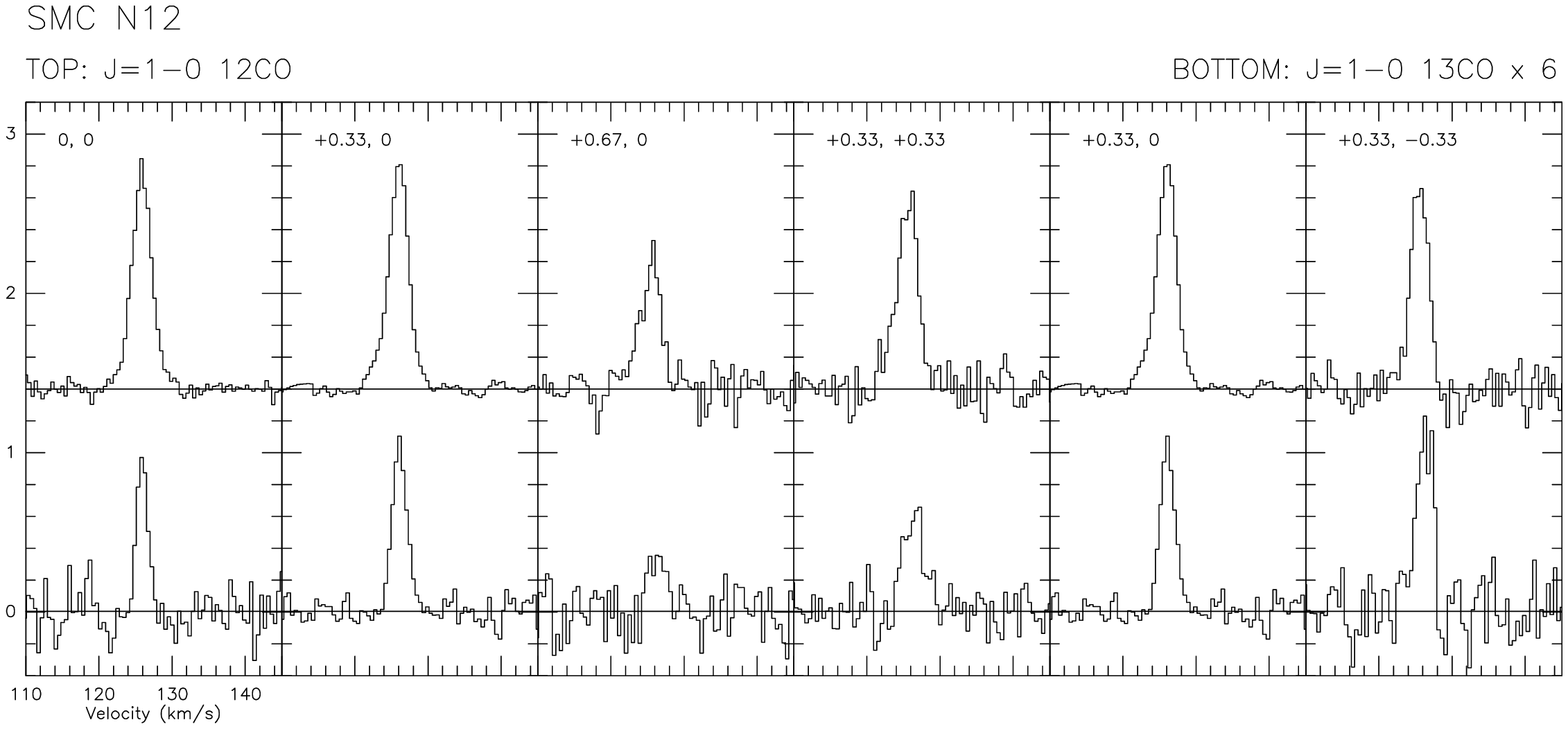}}}

\vspace{0.12cm}
\resizebox{8.9cm}{!}{\rotatebox{270}{\includegraphics*{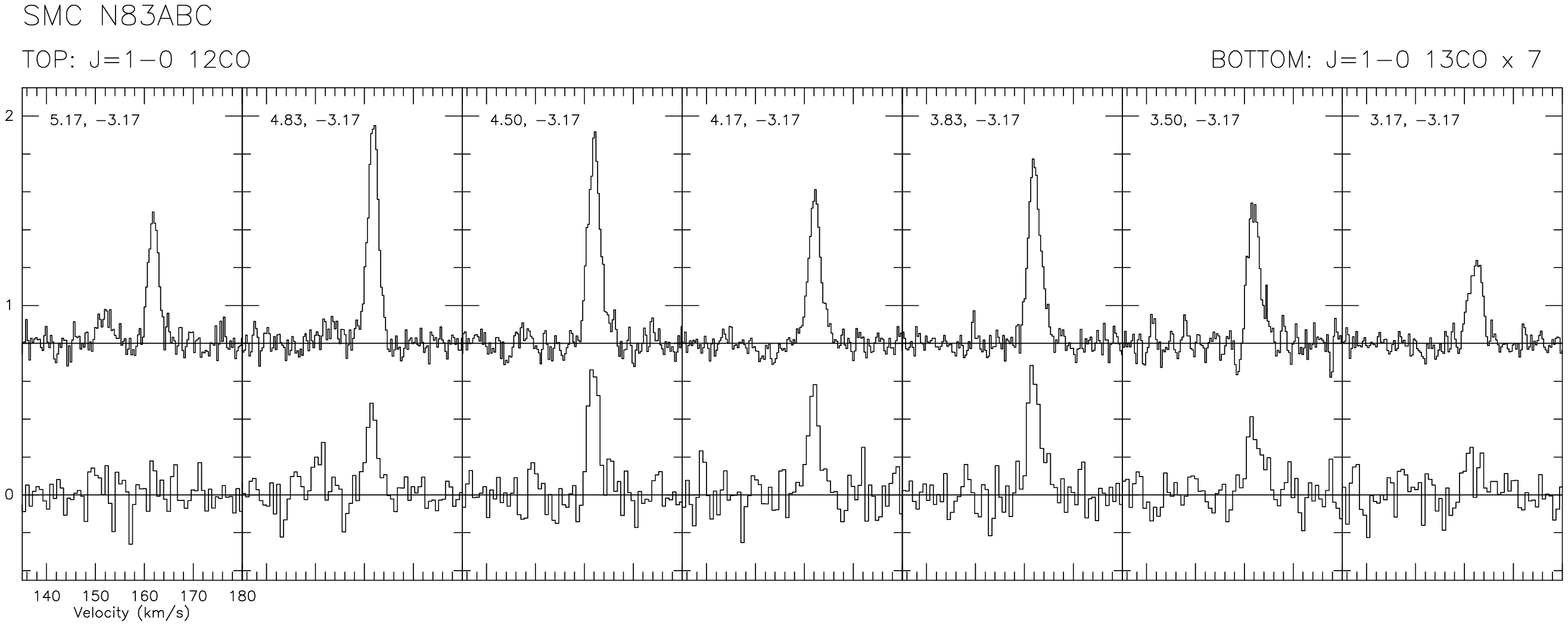}}}

\vspace{0.12cm}
\resizebox{8.9cm}{!}{\rotatebox{270}{\includegraphics*{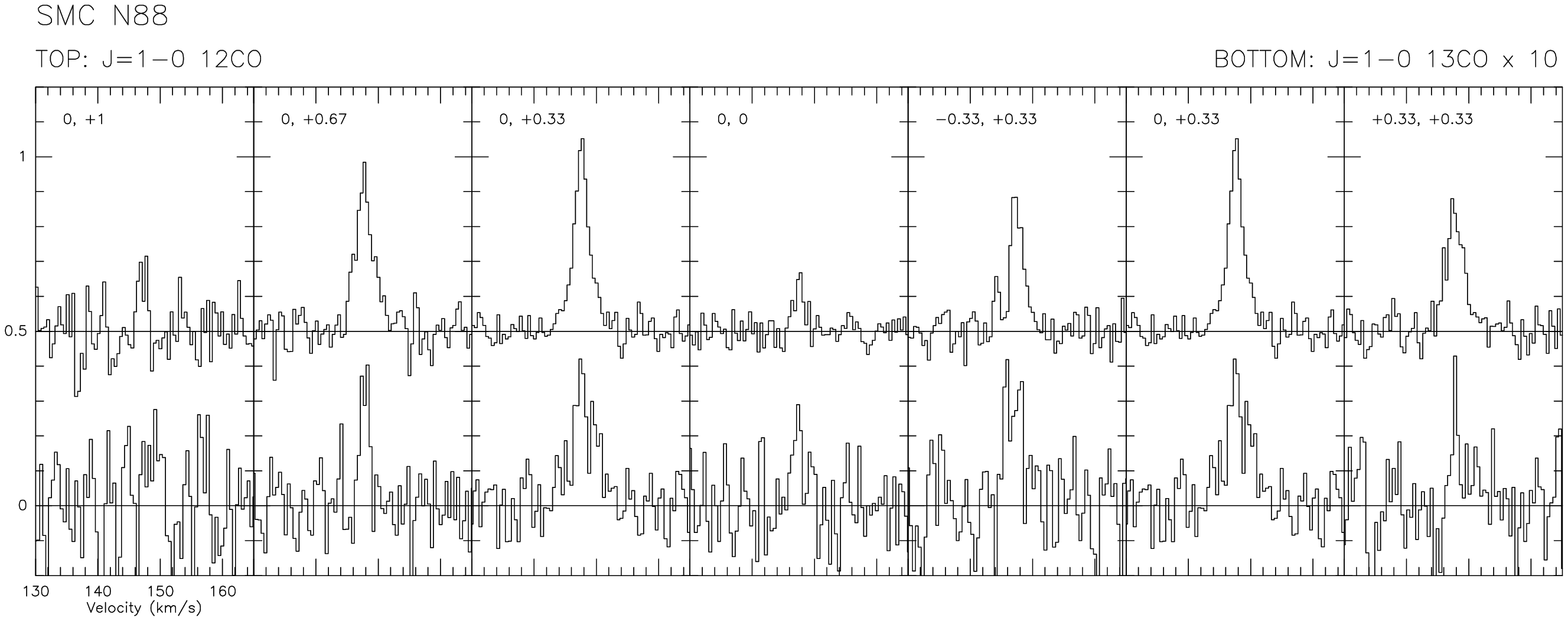}}}


\vspace{0.12cm}
\resizebox{8.9cm}{!}{\rotatebox{270}{\includegraphics*{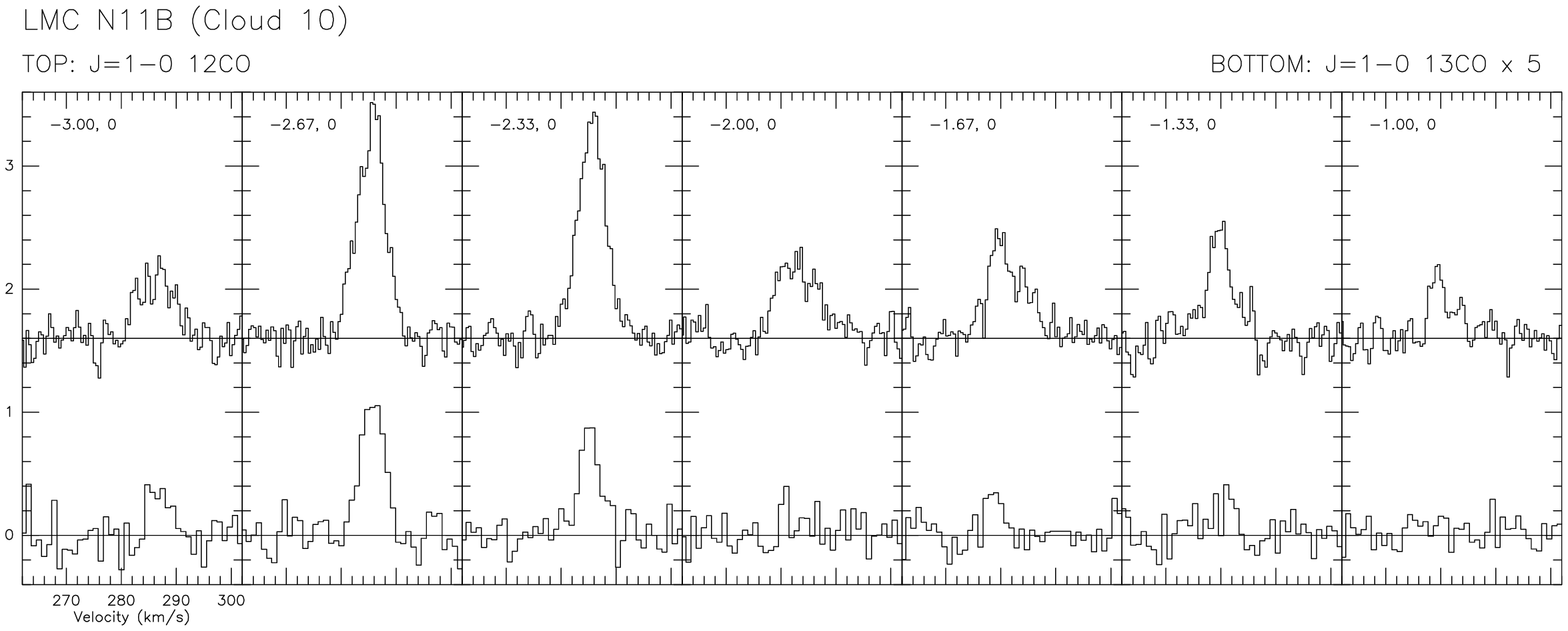}}}

\vspace{0.12cm}
\resizebox{8.9cm}{!}{\rotatebox{270}{\includegraphics*{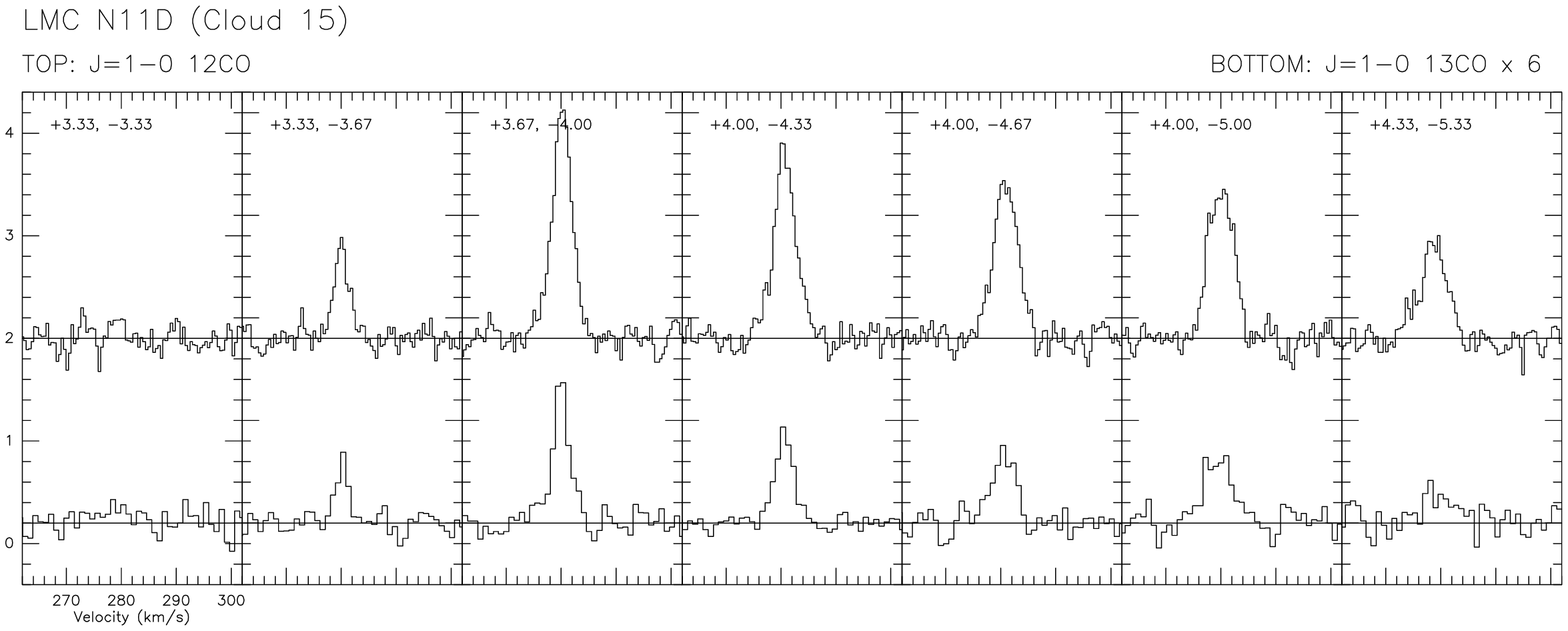}}}

\vspace{0.12cm}
\resizebox{8.9cm}{!}{\rotatebox{270}{\includegraphics*{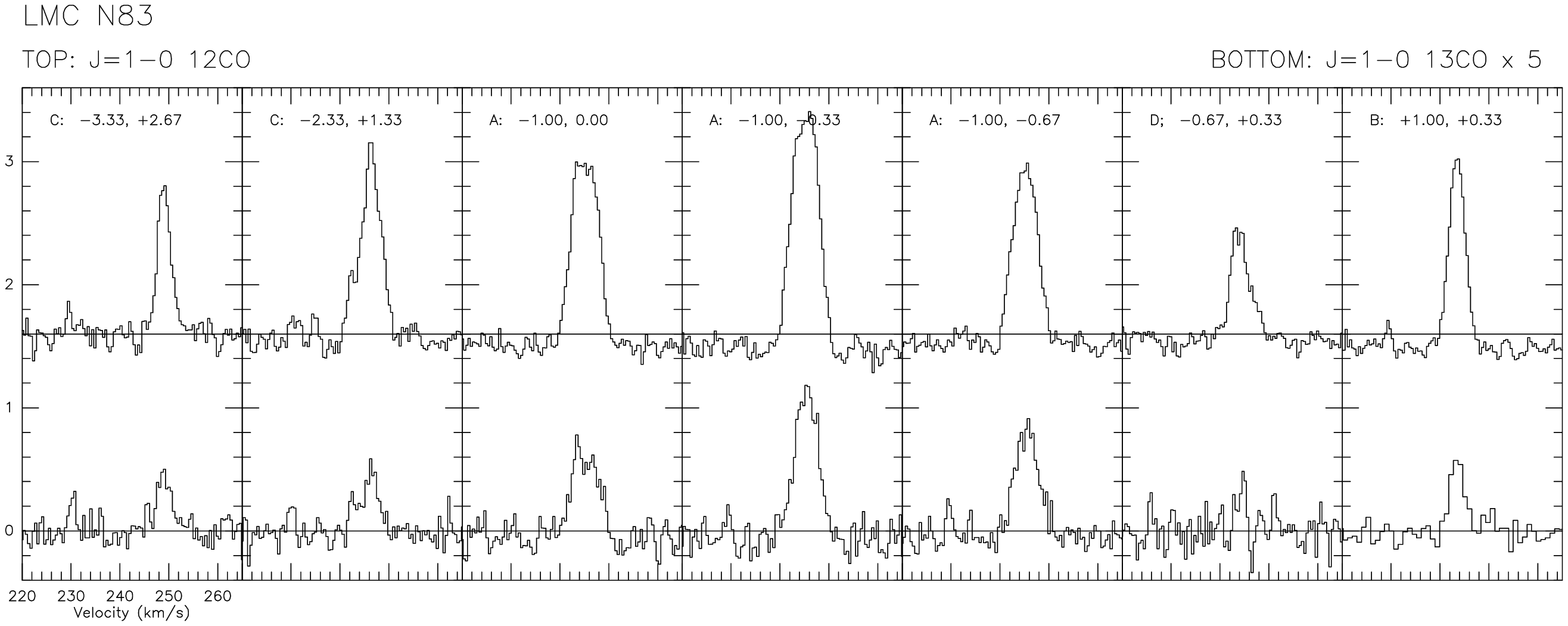}}}
\end{minipage}
\caption[]{Comparison of $\co$ and $\13co$ profiles in the clouds
associated with various HII regions. Temperature scales are in 
$T_{A}^{*} = 0.72 T_{mb}$. The offset positions refer to the integrated
emission maps as in Figs.~\ref{n12},\,\ref{n88} and \ref{n55}.
}
\label{prof}
\end{figure}

In 1988, a joint ESO-Swedish Key Programme was established on the
SEST to investigate the molecular gas in the Magellanic Clouds. 
The purpose of the Programme was twofold. First, it was intended to
establish the relation between CO emission and the much more
abundant molecular hydrogen gas it traces. Second, it intended to
map CO emission from individual molecular complexes and study
its relation to star formation. Finally, the Programme intended
publish a homogeneous set of molecular line data useful for further 
studies of the Magellanic Clouds. It was noted that the Magellanic 
Clouds allow investigation ofmolecular gas and cloud complexes under 
conditions of low metallicity and high radiation densities as compared 
to those found in the Solar Neighbourhood, and in fact different in 
the Large and the Small Magellanic Cloud.

The rationale of the programme was described in more detail by 
Israel et al. (1993; Paper I), who also presented the first
results from the Programme in the form of an extensive survey of CO
emission from (mostly far-infrared) sources in both the Large and
the Small Magellanic Cloud. This was followed by studies of the
major star-forming regions in the LMC (30 Doradus and N~159/N~160:
Johansson et al. 1998; N~11: Israel et al. 2003), by studies of
less active extended cloud complexes in the LMC (Kutner et al. 1997;
Garay et al. 2002), and by studies of relatively small molecular cloud 
complexes in the SMC (Rubio et al. 1993a, 1993b, 1996; Lequeux et al. 
1994). With ESO's discontinuation of the Key Programme concept, the 
observational programme was concluded in 1995, although the processing 
of data obtained has continued, as have observations of Magellanic Cloud 
objects independently of the Key Programme consortium (SMC-N~66: Rubio 
et al. 2000; LMC-N~159/N~160; SMC-N~83/N~84: Bolatto et al. 2000; 2003).
A study of star formation in LMC objects, based on a part of the Key 
Programme observations, was published by Caldwell $\&$ Kutner (1996). 
They found that, compared to the Milky way, LMC molecular clouds are 
less luminous in both the CO line and in the far-infrared continuum 
and that they are subject to significant massive star formation, 
irrespective of cloud (virial) mass. In this paper, we present the 
remaining part of the Key Programme observations, dealing with several 
molecular clouds associated with HII regions in both the Large and the 
Small Magellanic Cloud. 

\section{Observations}

\begin{table*}
\caption[]{SEST CO observations of Magellanic cloud Henize HII regions}
\begin{flushleft}
\begin{tabular}{lcccccccl}
\hline
\noalign{\smallskip}
Name$^{a}$ & DEM$^{b}$ & LI-LMC$^{c}$ & SMC-J$^{d}$ & NGC & \multicolumn{2}{c}{Map Center$^{f}$} & 
Map Size ($'$) & Reference \\
  &  & LI-SMC  & LMC-B$^{e}$ & & $\alpha$(1950.0) & $\delta$(1950.0) & $\Delta\alpha \times \Delta\delta$ & \\
\noalign{\smallskip}
\hline
\noalign{\smallskip}
\noalign{\smallskip}
{\bf SMC}&        &            &               &      &            &           &               &\\
\noalign{\smallskip}
N12A     &  23    &  36        & 004633-730604 & ---  & 00:44:54.0 & -73:22:29 & $4\times3$    & TP, 7 \\
N13A,B   &  16    &  29        & 004523-732250 & ---  & 00:44:01.3 & -73:39:10 & $7.3\times4.7$& 8\\
N15      &  28    &  ---       & ---           & ---  &            &           &               & 8\\
N16      &  21    &  35        & 004619-732324 & 248  &            &           &               & 8\\
N22      &  37    &  45        & 004834-731509 & ---  & 00:46:15.9 & -73:29:50 & $4.3\times10$ & 8 \\
N25/N26  &  38    &  45        & 004834-731509 & ---  &            &           &               & 8\\
N27      &  40    &  49        & 004823-730557 & ---  & 00:46:28.7 & -73:21:32 & $2.6\times2.6$& TP, 7\\
N66A-D   & 103    & 131,135,137& 0059**-7210** & 346  & 00:59:20.0 & -72:10:0  & $4\times4$    & 7, 9 \\
N83A-C   & 147-148& 199        & 011416-731549 & 456  & 01:12:43.4 & -73:32:03 & $3\times3.5$  & TP, 10\\
N84C     & 149    & 200        & 011416-731549 & 456  &            &           &               & TP, 10\\
N84B,D   & 152    & 202        & ---           & 456  & 01:13:23.0 & -73:36:33 & $1.6\times1.6$& TP, 10\\
N88      & 161    & 215        & 012408-730905 & ---  & 01:22:54.1 & -73:24:15 & $1.4\times2.0$& TP, 7\\
\noalign{\smallskip}
{\bf LMC}&        &            &               &      &            &           &               &\\
\noalign{\smallskip}
N4A,B    &   8    & 102        & 0452-6700     & 1714 & 04:52:02.9 & -67:00:05 & $3\times2.5$  & 1\\
N83A-D   &  22    & 148,173,193& 0454-6916     & 1737 & 04:54:12
.1 & -69:15:00 & $7\times5$    & TP\\
N11A-J   &  34    & 190,192,195&   ---         & 1760 & 04:56:57.3 & -66:27:00 & $28\times36$  & 2\\
         &        & 205,214,217& 0456-6629/6636& 1763 &            &           &               & \\
         &        & 226,229,243& 0457-6632     & 1769 &            &           &               & \\
         &  41    & 248,251,   & 0458-6626     & 1773 &            &           &               & \\
         &        & 266,268,271& 0458-6616     & ---  &            &           &               & \\
N55A     & 228    & 1268,1273  & 0532-6629     & ---  & 05:32:31.6 & -66:28:00 & $4.5\times4$  & TP\\
N57A/E   & 229,231& 1261,1274  & 0532-6743     & 2014 & 05:32:30.0 & -67:43:05 & $12\times12$  & TP\\
N59A-C   & 241    & 1367,1392  & 0535-6736    & 2032/5& 05:36:00.0 & -67:36:05 & $12\times12$  & TP\\
N157A,B  & 263    & 1469       & 0538-6911     & 2060 & 05:38:09.5 & -69:07:00 & $27\times26$  & 3\\
N159A-L  & 271-272& 1501,1518  & 0540-6946     & 2079 & 05:40:18.2 & -69:42:30 & $13\times25$  & 3, 4\\
N160A-F  & 283-284& 1503,1549  & 0540-6940     & 2080 &            &           &               & 3, 4\\
N158C,D  & 269    & 1490       & 0539-6931     & 2074 &            &           &               & 3, \\
N214A-C,E& 274,278,293& 1505,1521,1577& 0540-7111& 2103& 05:40:35.8& -71:11:00 & $9\times10$   & 5\\
N171A,B  & 267    & 1486       &  ---          & ---  & 05:40:41.8 & -70:22:00 & $13\times40$  & 5\\
N176     & 280    & 1541       &  ---          & ---  &            &           &               & 5\\
N167     & 307    & 1633       &  ---          & ---  & 05:45:24.5 & -69:26:00 & $19\times15$  & 6\\
N72      & 304    & 1602       & 0543-6918     & ---  &            &           &               & 6\\
N169A-C  & 312-314& 1696       & 0546-6934     & ---  & 05:46:11.6 & -69:38:00 & $16\times8$   & 6\\
\noalign{\smallskip}
\hline
\end{tabular}
\end{flushleft}
Notes: 
$^{a}$ Optical designation by Henize (1956); 
$^{b}$ Optical designation by Davies et al. (1976);
$^{c}$ IRAS designation by Schwering $\&$ Israel (1990). 
$^{d}$ ATCA radio continuum designation by Filipovic et al. (2002).
$^{e}$ Parkes radio continuum designation by Filipovic et al. (1996).
$^{f}$ Actual center of map, not to be confused with (0,0) map reference. \\
References: TP: This Paper
            1.: Heydari-Malayeri $\&$ Lecavelier des Etangs (1994);
            2.: Israel et al. (2003);
            3.: Johansson et al. (1998);
            4.: Bolatto et al. (2000);
	    5.: Kutner et al. (1997);
            6 : Garay et al. (2002);
	    7.: Rubio et al. (1996);
            8.: Rubio et al. (1993);
	    9.: Rubio et al. (2000);
	   10.: Bolatto et al. (2003);
\end{table*}

\begin{table*}
\caption[]{Magellanic Cloud HII regions observed in CO emission}
\begin{flushleft}
\begin{tabular}{lccccccrrr}
\hline
\noalign{\smallskip}
Name   & \multicolumn{5}{c}{Peak CO Parameters} & \multicolumn{4}{c}{Cloud CO Parameters}  \\
       & T$_{\rm mb}$ & $\Delta$V & $V_{\rm LSR}$ & $I_{\rm CO}$ & $I_{12}/I_{13}$ & Radius$^{a}$ & $L_{\rm CO}$ & $M_{vir}$ & $X$ \\
       &  (K)        &\multicolumn{2}{c}{($\kms$)} &  ($\kkms$)  &    & (pc) & ($\kkms$pc$^{2}$) & 10$^{4} \Msun$ & ($10^{20} \h2$ cm$^{-2} (\kkms)^{-1})$ \\
\noalign{\smallskip}
\hline
\noalign{\smallskip}
\noalign{\smallskip}
{\bf SMC} &     &     &       &      &            &    &       &      & \\
\noalign{\smallskip}
N12       & 2.0 & 3.4 & 126.0 &  8.2 &  11$^{d}$  &  7 &  3970 &  1.7 &  2.0\\
N27       & 1.7 & 6.6 & 113.2 & 11.8 &  16$^{d}$  & 10 &  8610 &  9.2 &  5.3\\
N83A-C$^{b}$& 1.1 & 3.1 & 161.7 & 5.4 & 10$^{e}$  & 14 &  4015 &  2.8 &  3.5\\
N84C      & 1.0 & 4.3 & 160.7 &  3.3 &  ---       &  5 &  1230 &  1.9 &  7.5\\
N84B,D    & 0.8 & 2.6 & 168.1 &  2.2 &  ---       &  5 &   880 &  0.6 &  3.5\\
N88       & 0.7 & 2.9 & 147.7 &  2.5 &  13$^{e}$  &  5 &   530 &  0.8 &  7.1\\
\noalign{\smallskip}
{\bf LMC} &     &     &       &      &            &    &       &      & \\
\noalign{\smallskip}
N55       & 3.5 & 4.8 & 289.0 & 16.7 &  12$^{c}$  &  9 &  6235 &  4.4 &  3.3\\
N57A1     & 1.2 & 3.2 & 287.8 &  4.3 &  13$^{c}$  &  4 &   715 &  0.9 &  5.7\\
N57A2     & 2.9 & 4.1 & 285.7 & 12.6 &  ---       & 13 &  3760 &  4.6 &  5.8\\
N57A3     & 3.1 & 2.8 & 284.5 &  9.1 &  ---       & 15 &  3225 &  2.5 &  3.8\\
N57A4     & 1.0 & 5.1 & 284.9 &  5.3 &  ---       &  6 &  1690 &  3.3 &  9.7\\
N57A5     & 1.3 & 3.0 & 286.3 &  5.1 &  ---       &  8 &  1290 &  1.5 &  5.6\\
N59A1     & 1.4 & 6.0 & 285.2 &  9.1 & \raisebox{-1ex}{12$^{c}$} &  7 & 5570 &  5.3 &  4.5\\
N59A2     & 1.8 & 2.9 & 279.9 &  5.3 &            &  7 &  3260 &  1.2 &  1.8\\
N59A3     & 2.2 & 3.7 & 282.4 &  8.1 &  ---       &  6 &  3035 &  1.7 &  2.7\\
N59A4     & 0.9 & 5.0 & 282.6 &  6.6 &  ---       & 16 & 15970 &  8.1 &  2.4\\
N83A      & 2.9 & 6.7 & 245.2 & 19.4 &   9$^{e}$  &  6 &  5370 &  5.7 &  5.3\\
N83B      & 2.2 & 4.1 & 243.5 &  9.7 &  11$^{e}$  &  4 &  2510 &  1.2 &  2.5\\
N83C$^{b}$& 2.2 & 5.3 & 246.3 & 12.8 &  14$^{e}$  & 12 & 10920 &  7.1 &  3.1\\
N83D      & 1.3 & 3.1 & 243.7 &  6.1 &  16$^{e}$  &$<$2&  530 &$<$0.4&$<$3.6\\
\noalign{\smallskip}
\hline
\end{tabular}
\end{flushleft}
Notes: 
$^{a}$ Corrected for finite beamwidth. 
$^{b}$ Complex Source.
$^{c}$ Israel et al. (1993)
$^{d}$ Rubio et al. (1996)
$^{e}$ This Paper
\end{table*}

\begin{figure*}[t]
\unitlength1cm
\begin{minipage}[t]{5.9cm}
\resizebox{6.cm}{!}{\rotatebox{270}{\includegraphics*{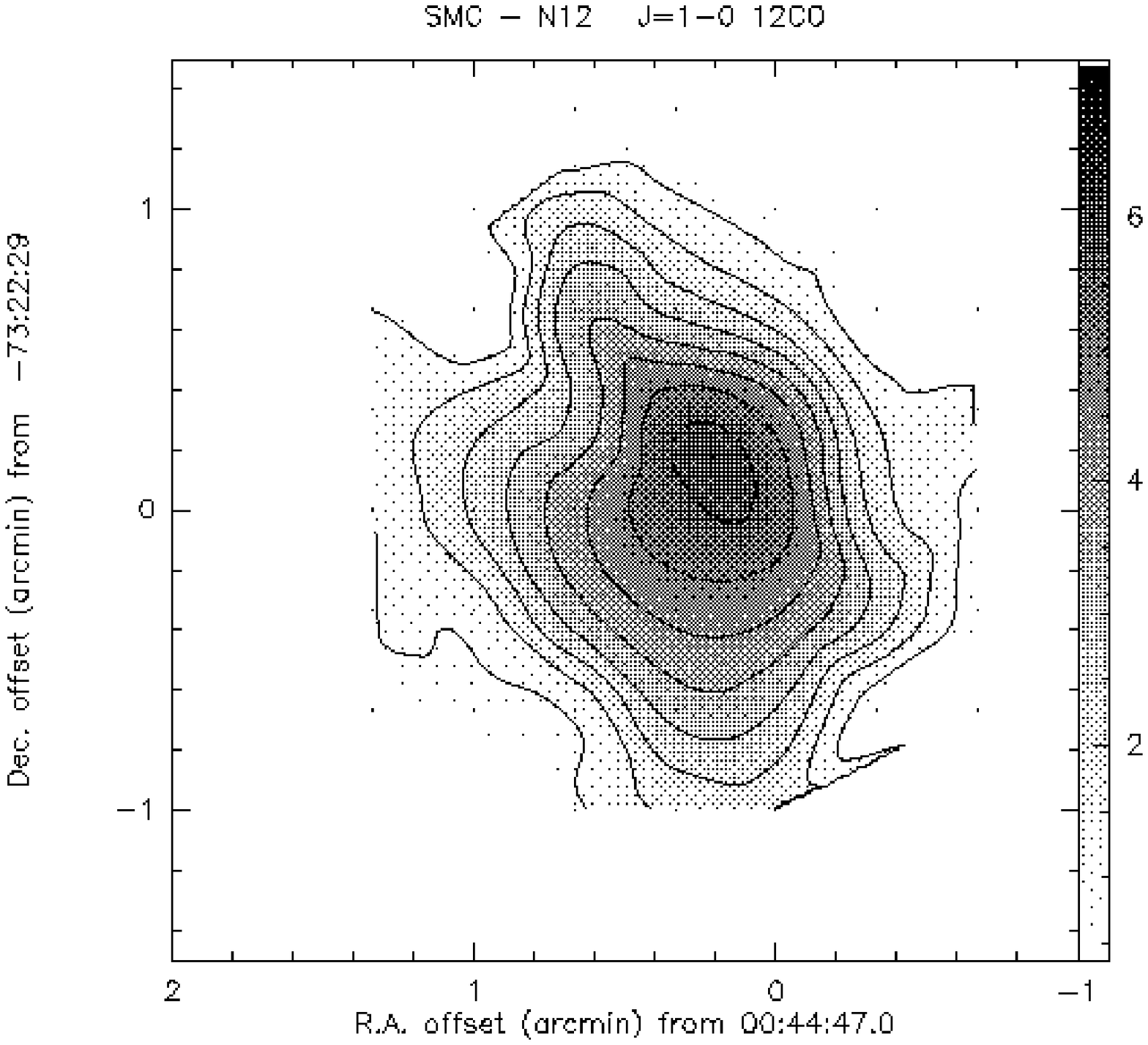}}}
\resizebox{6.cm}{!}{\rotatebox{270}{\includegraphics*{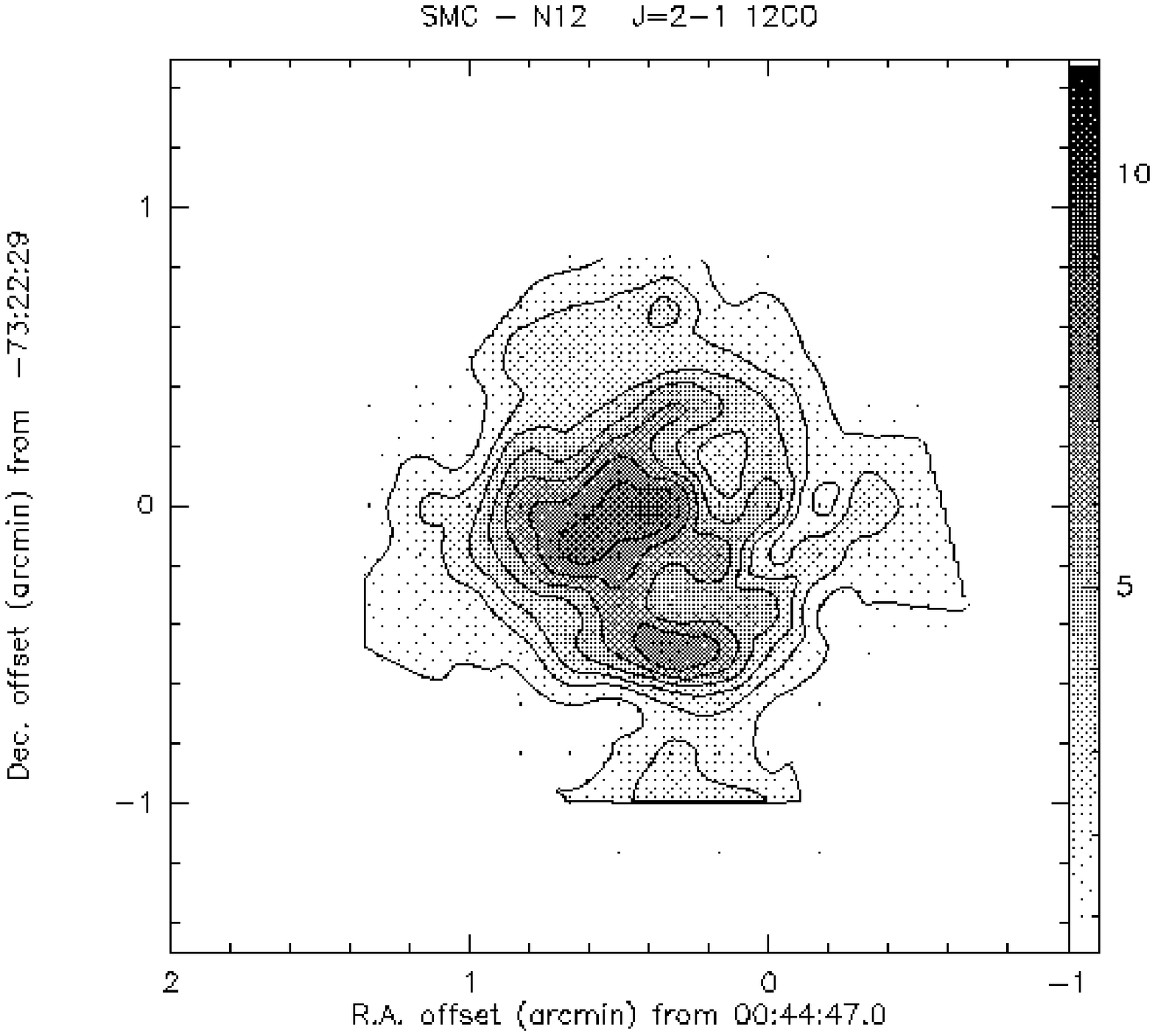}}}
\resizebox{6.cm}{!}{\rotatebox{270}{\includegraphics*{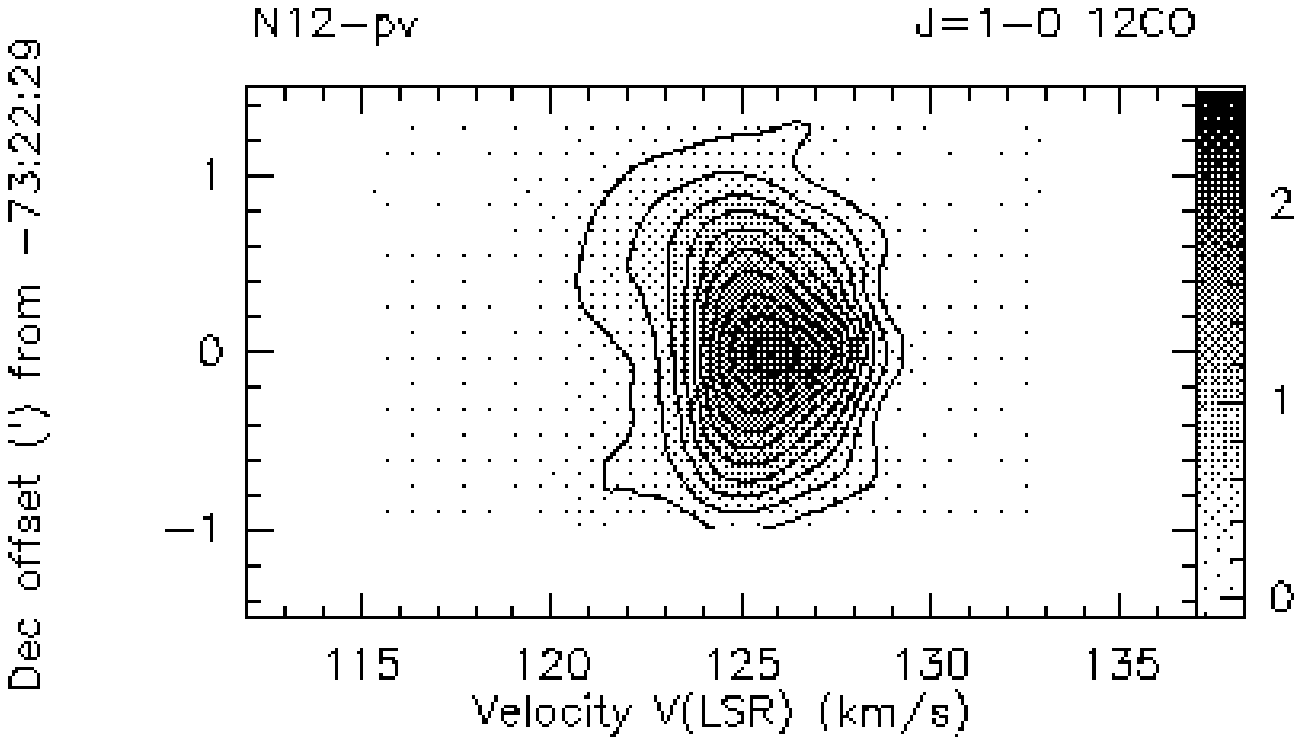}}}
\resizebox{6.cm}{!}{\rotatebox{270}{\includegraphics*{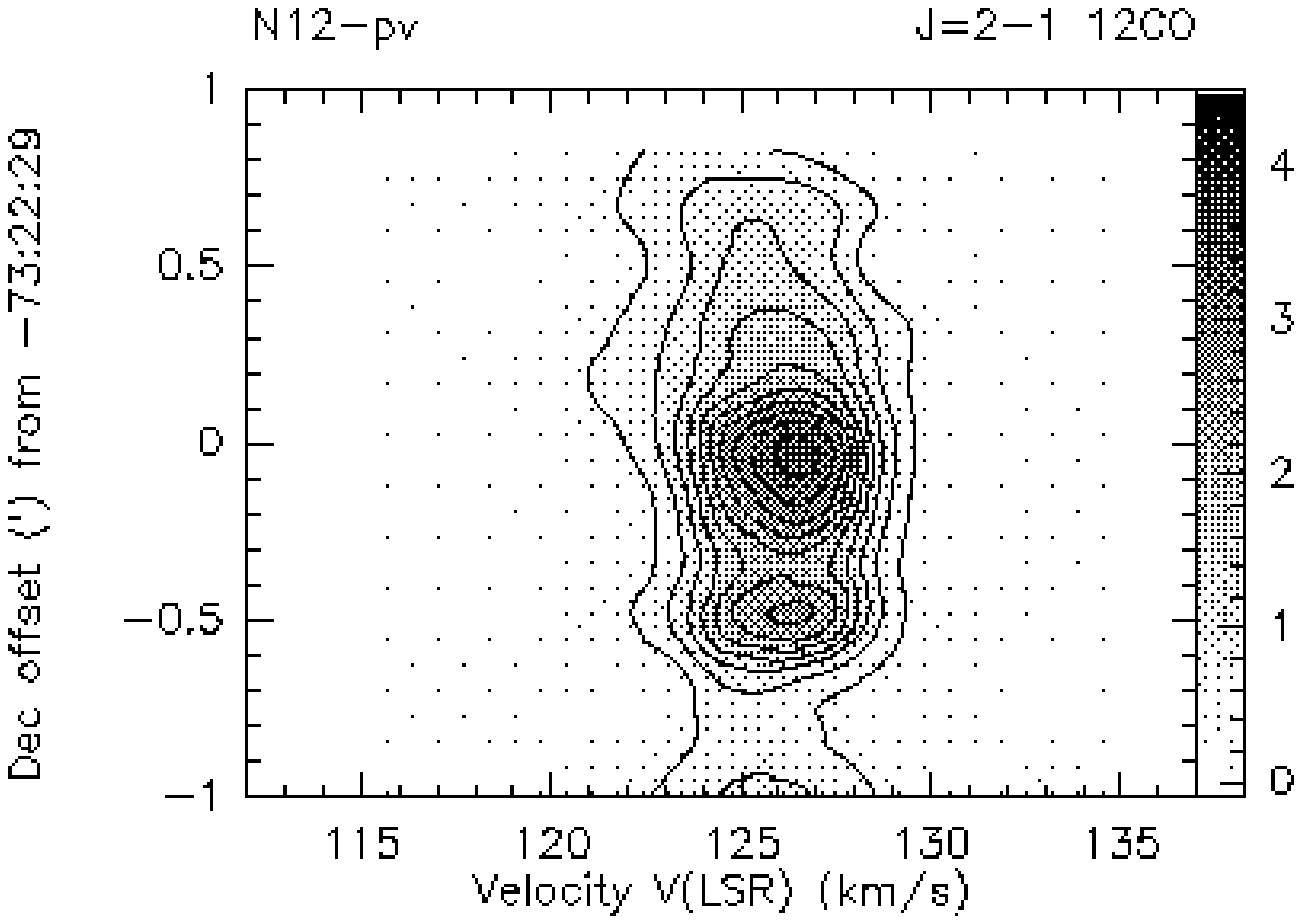}}}
\end{minipage}
\hfill
\begin{minipage}[t]{5.9cm}
\resizebox{6.cm}{!}{\rotatebox{270}{\includegraphics*{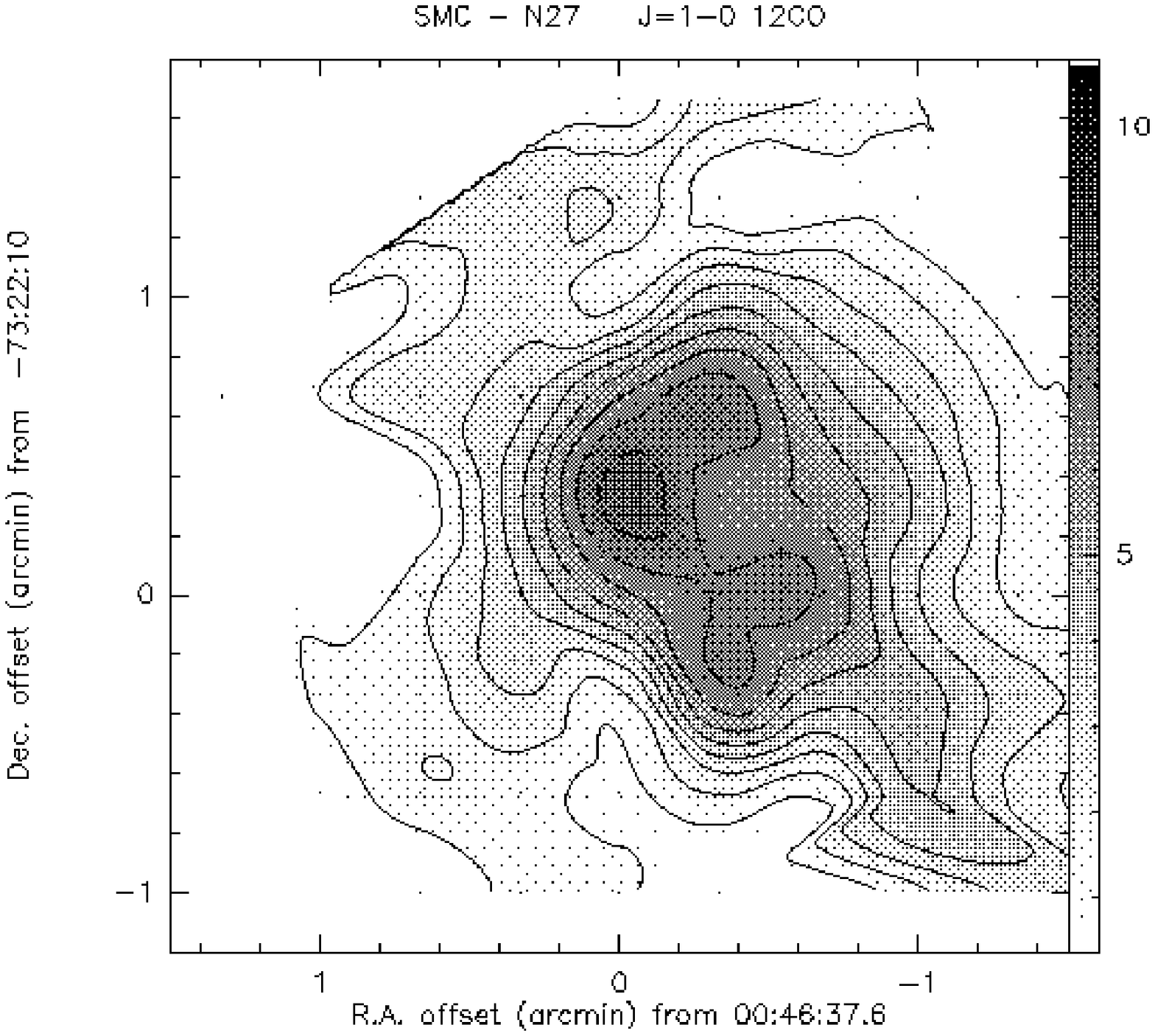}}}
\resizebox{6.cm}{!}{\rotatebox{270}{\includegraphics*{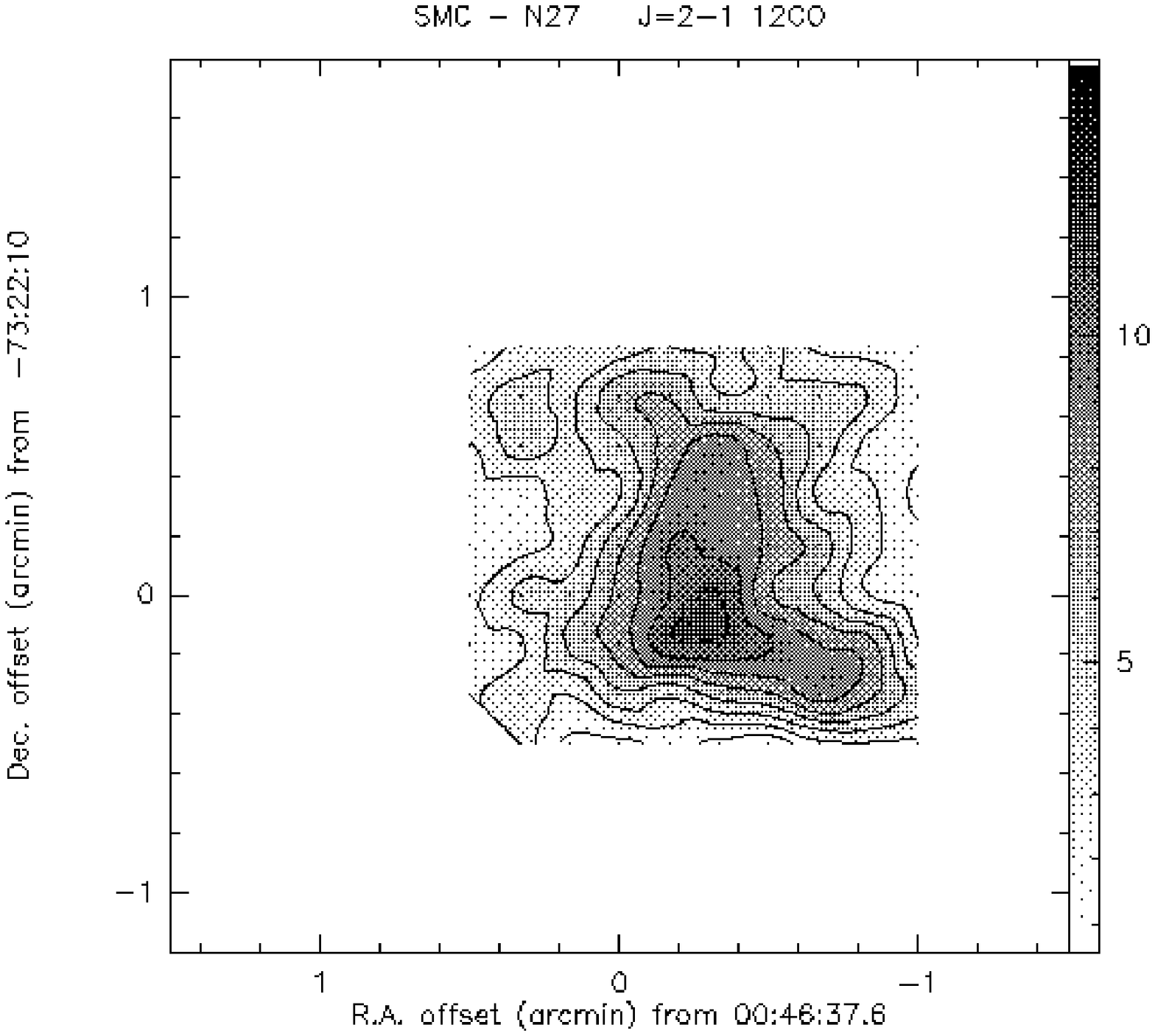}}}
\resizebox{6.cm}{!}{\rotatebox{270}{\includegraphics*{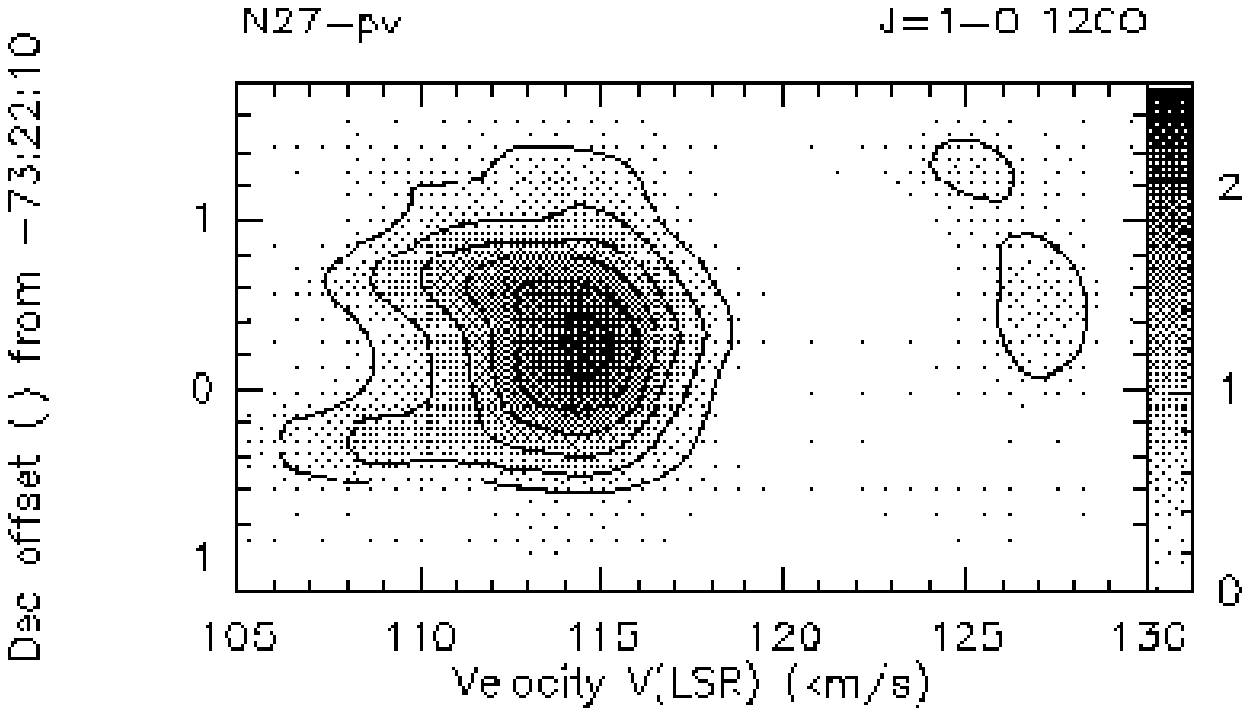}}}
\resizebox{6.cm}{!}{\rotatebox{270}{\includegraphics*{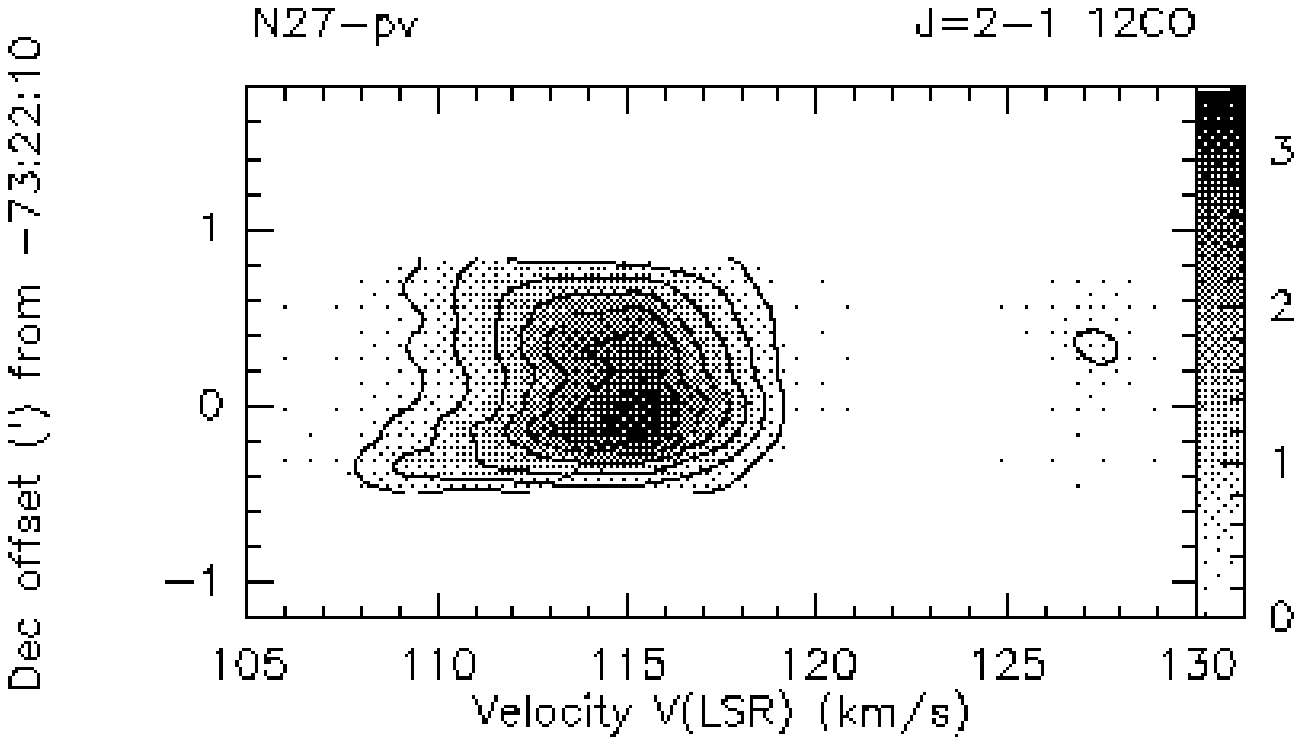}}}
\end{minipage}
\hfill
\begin{minipage}[t]{5.9cm}
\resizebox{6.cm}{!}{\rotatebox{270}{\includegraphics*{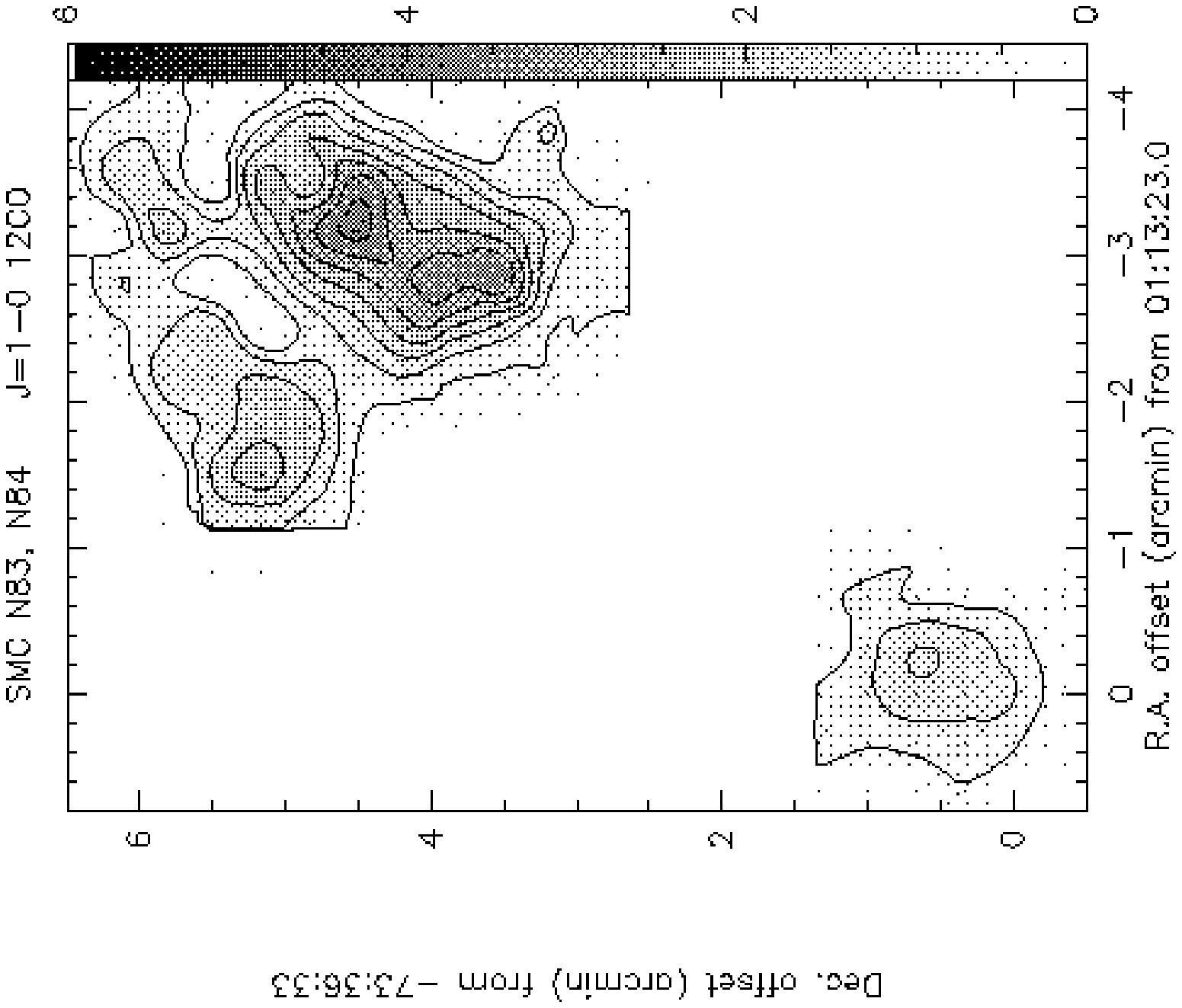}}}
\resizebox{6.cm}{!}{\rotatebox{270}{\includegraphics*{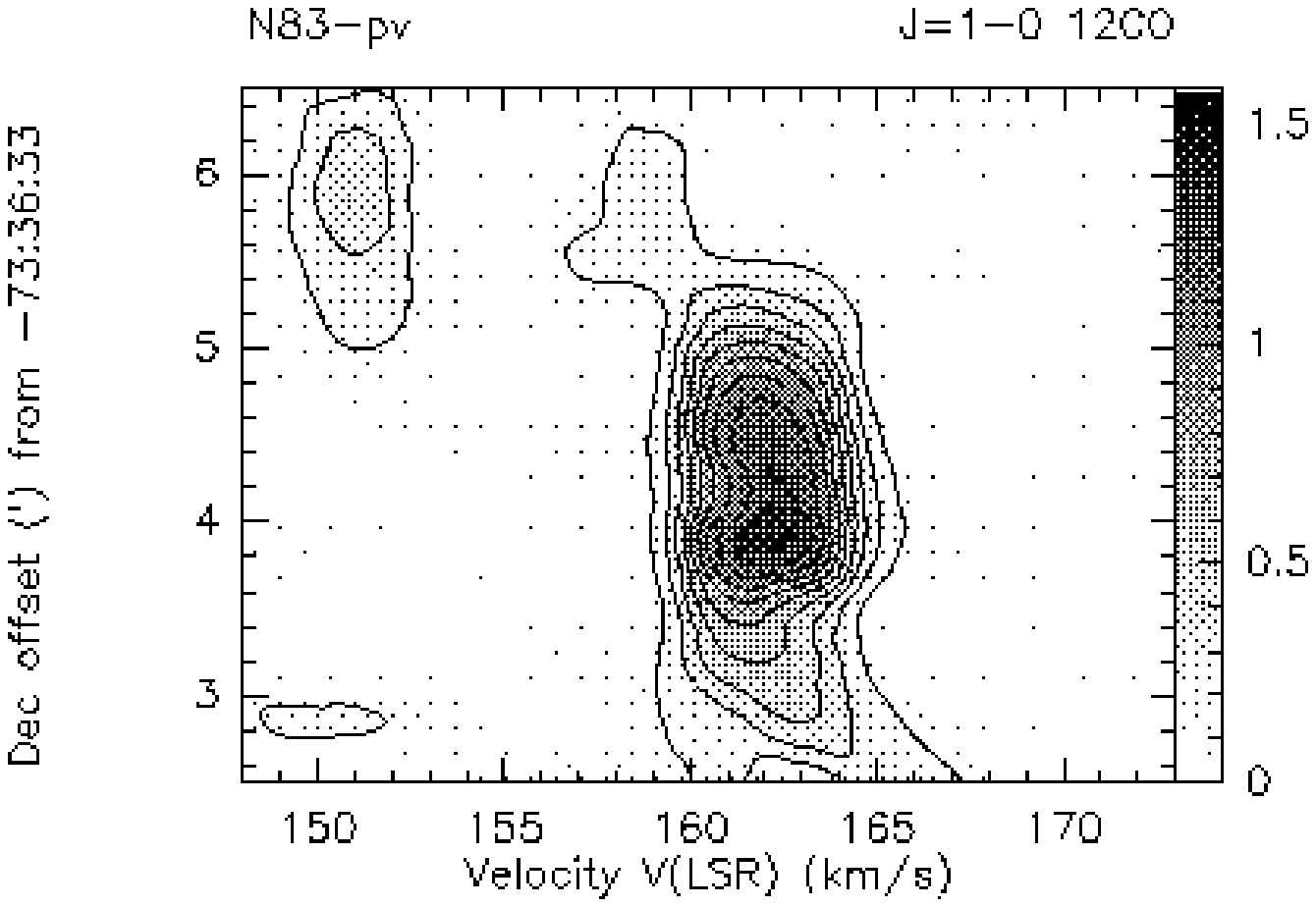}}}
\resizebox{6.cm}{!}{\rotatebox{270}{\includegraphics*{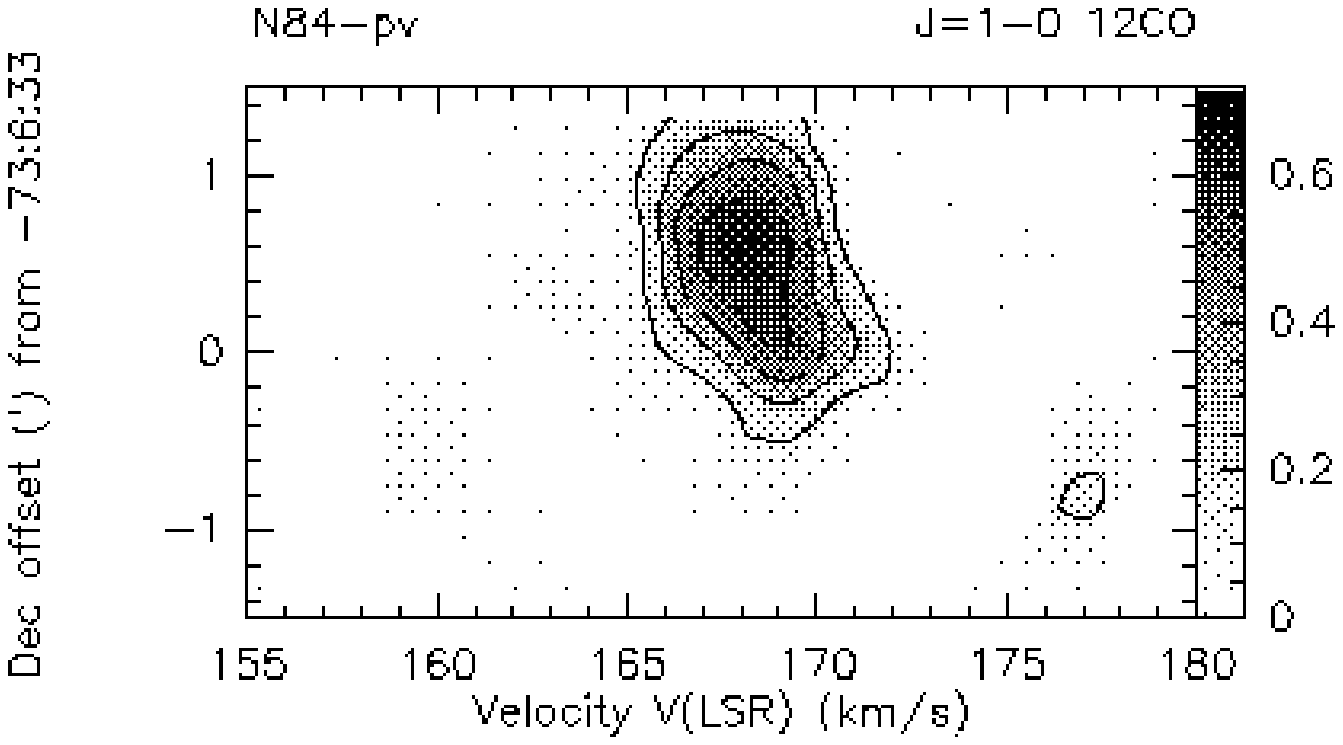}}}
\end{minipage}
\caption[]{
Maps of CO clouds associated with SMC HII regions N~12, N~27, N~83 and 
N~84. Linear contours are at multiples of $\int$ $T_{mb}$d$V$ = 1 $\kkms$ 
for N~12 (both transitions), 1.25 $\kkms$ ($J$=1-0) and 1 $\kkms$ ($J$=2-1) 
for N~27 and 0.75 $\kkms$ for N~83 and N~84. For all four objects, 
position-velocity cuts are in declination. Linear contours are at 
multiples of $T_{mb}$ = 0.3 K for N~12, at 0.40 K and 0.75 K respectively 
for N~27 and at 0.20 K for N~83 and N~84. Gray scales are labelled in 
$T_{A}^{*}$ rather than $T_{mb}$.
}
\label{n12}
\end{figure*} 

\begin{figure*}[t]
\unitlength1cm
\begin{minipage}[t]{5.9cm}
\resizebox{6.cm}{!}{\rotatebox{270}{\includegraphics*{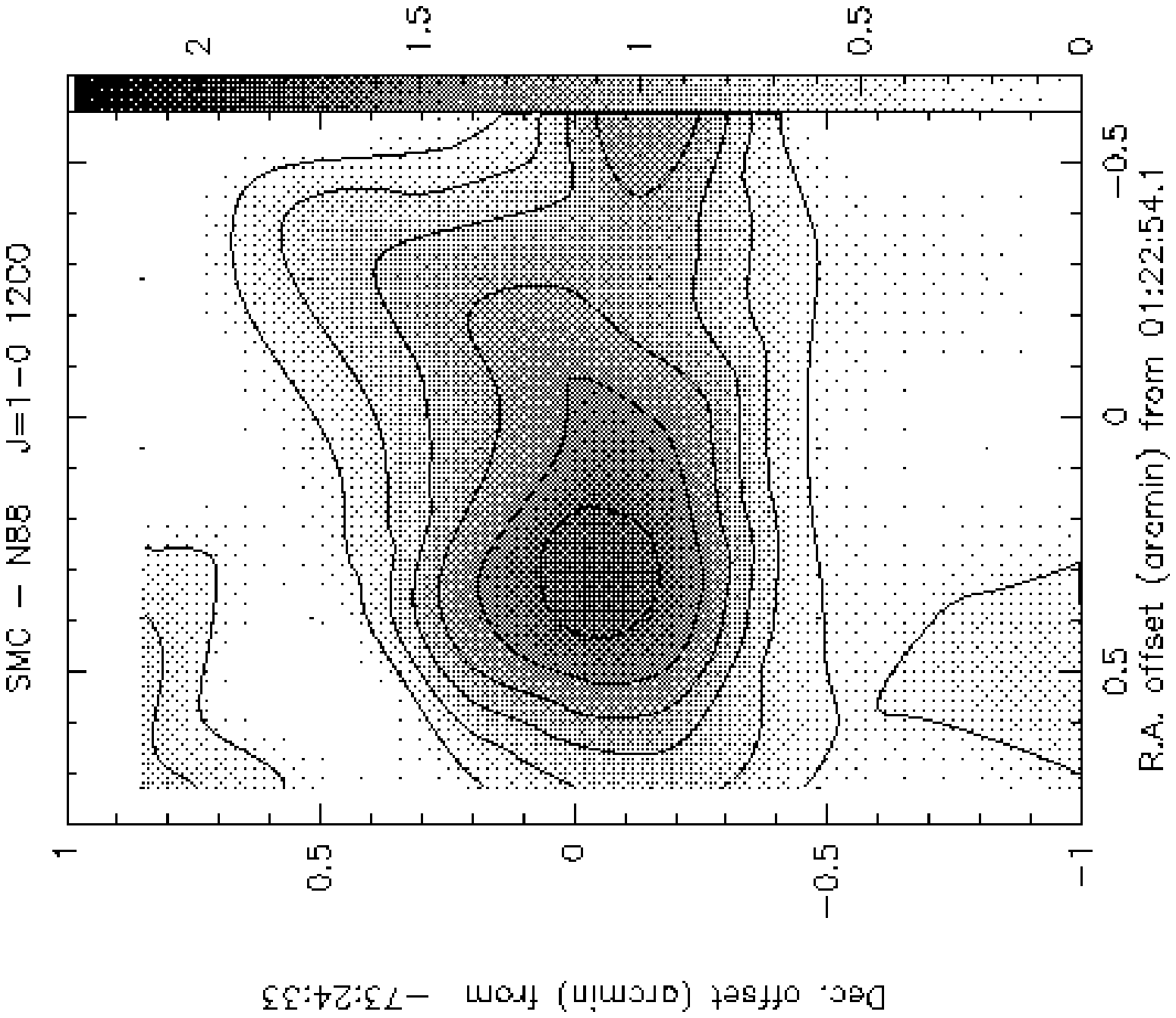}}}
\end{minipage}
\hfill
\begin{minipage}[t]{5.9cm}
\resizebox{6.cm}{!}{\rotatebox{270}{\includegraphics*{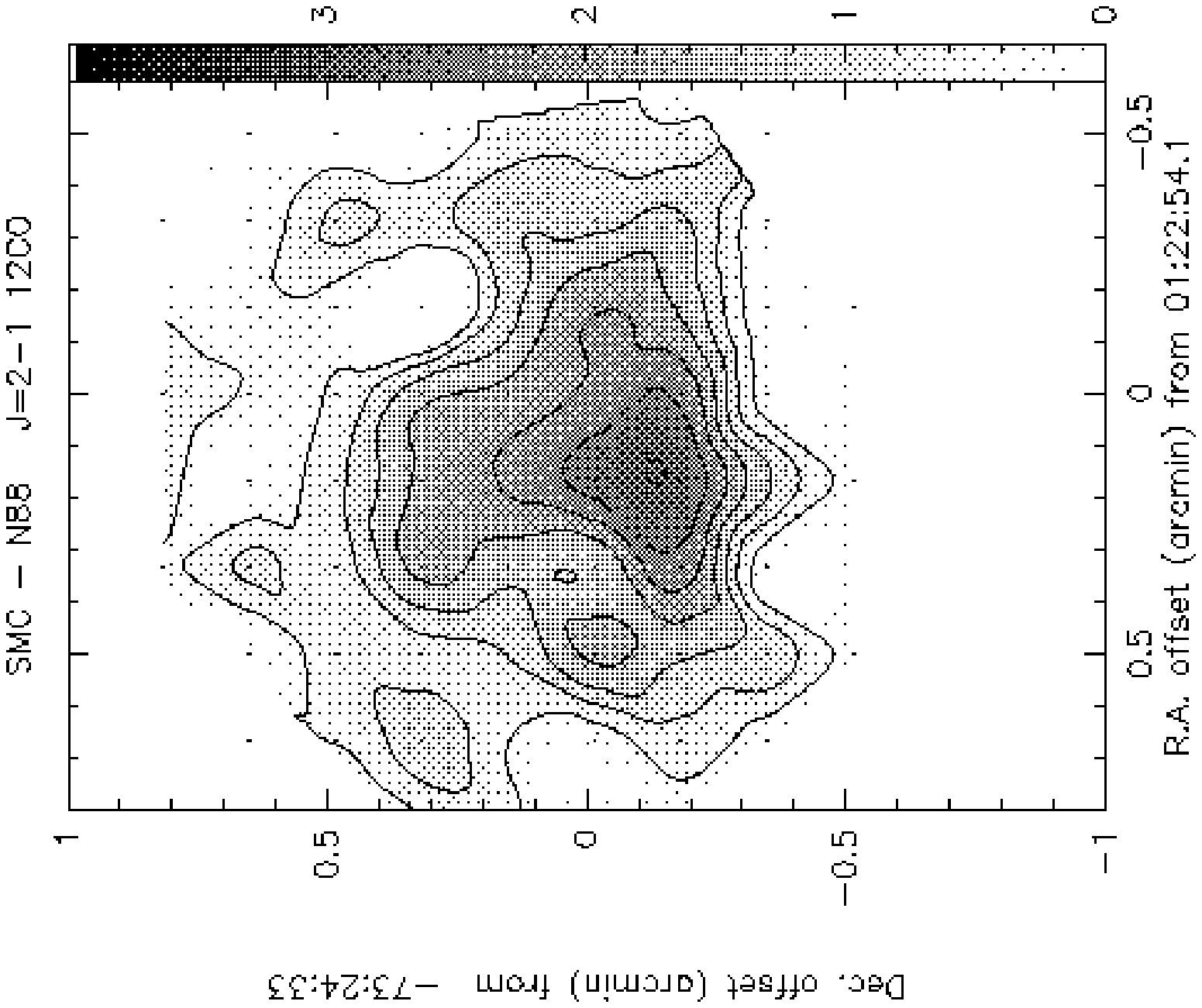}}}
\end{minipage}
\hfill
\begin{minipage}[t]{5.9cm}
\resizebox{6.cm}{!}{\rotatebox{270}{\includegraphics*{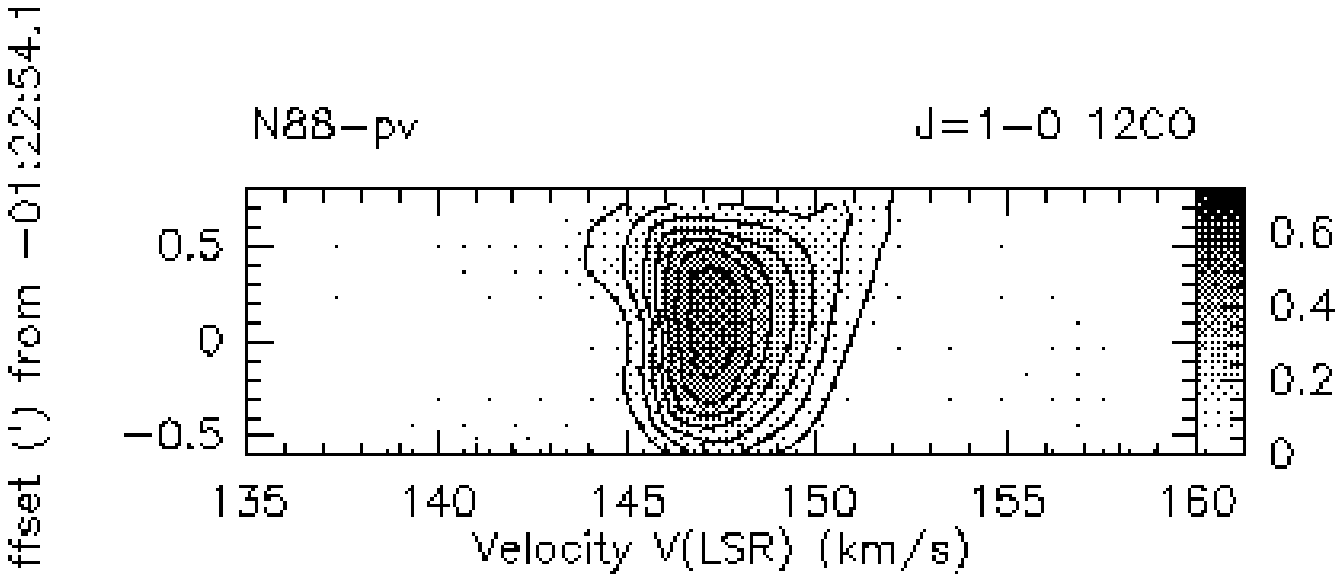}}}
\resizebox{6.cm}{!}{\rotatebox{270}{\includegraphics*{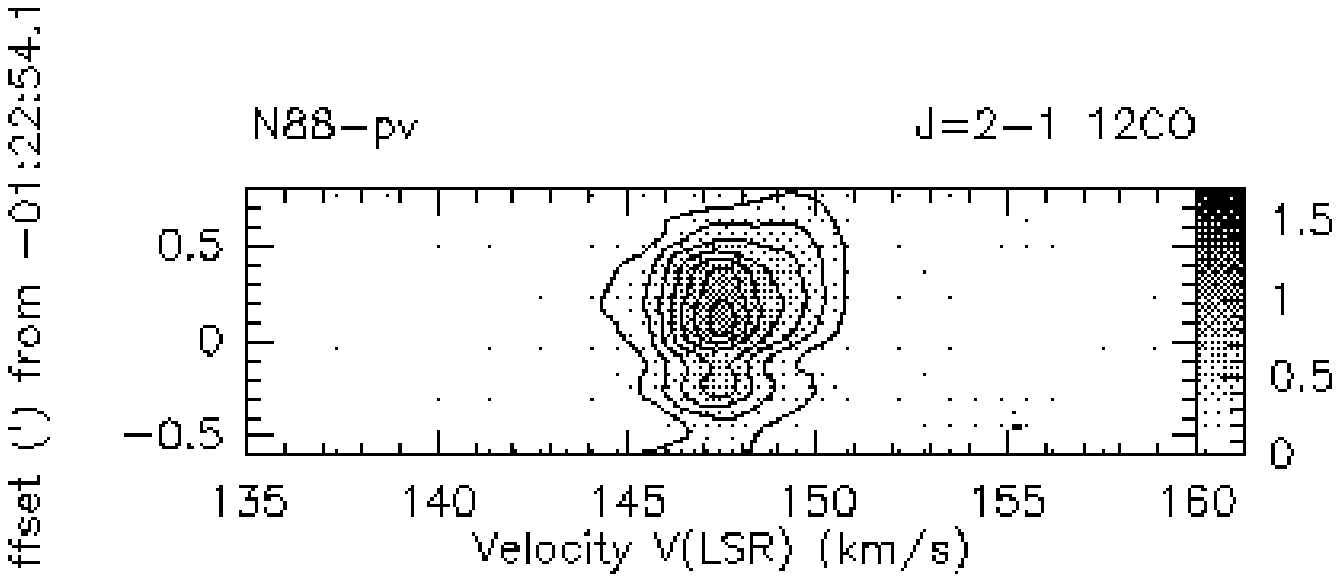}}}
\end{minipage}
\caption[]{
Maps of CO cloud associated with SMC HII region N~88. Linear contours are 
at multiples of $\int$ $T_{mb}$d$V$ = 0.40 $\kkms$ ($J$=1-0) and 0.75 
$\kkms$ ($J$=2-1). Position velocity cuts are in right ascension. Linear 
contours are at multiples of $T_{mb}$ = 0.10 K and 0.25 K respectively. 
Gray scales are labelled in $T_{A}^{*}$ rather than $T_{mb}$.
}
\label{n88}
\end{figure*} 

\begin{figure*}
\unitlength1cm
\begin{minipage}[t]{5.9cm}
\resizebox{6.cm}{!}{\rotatebox{270}{\includegraphics*{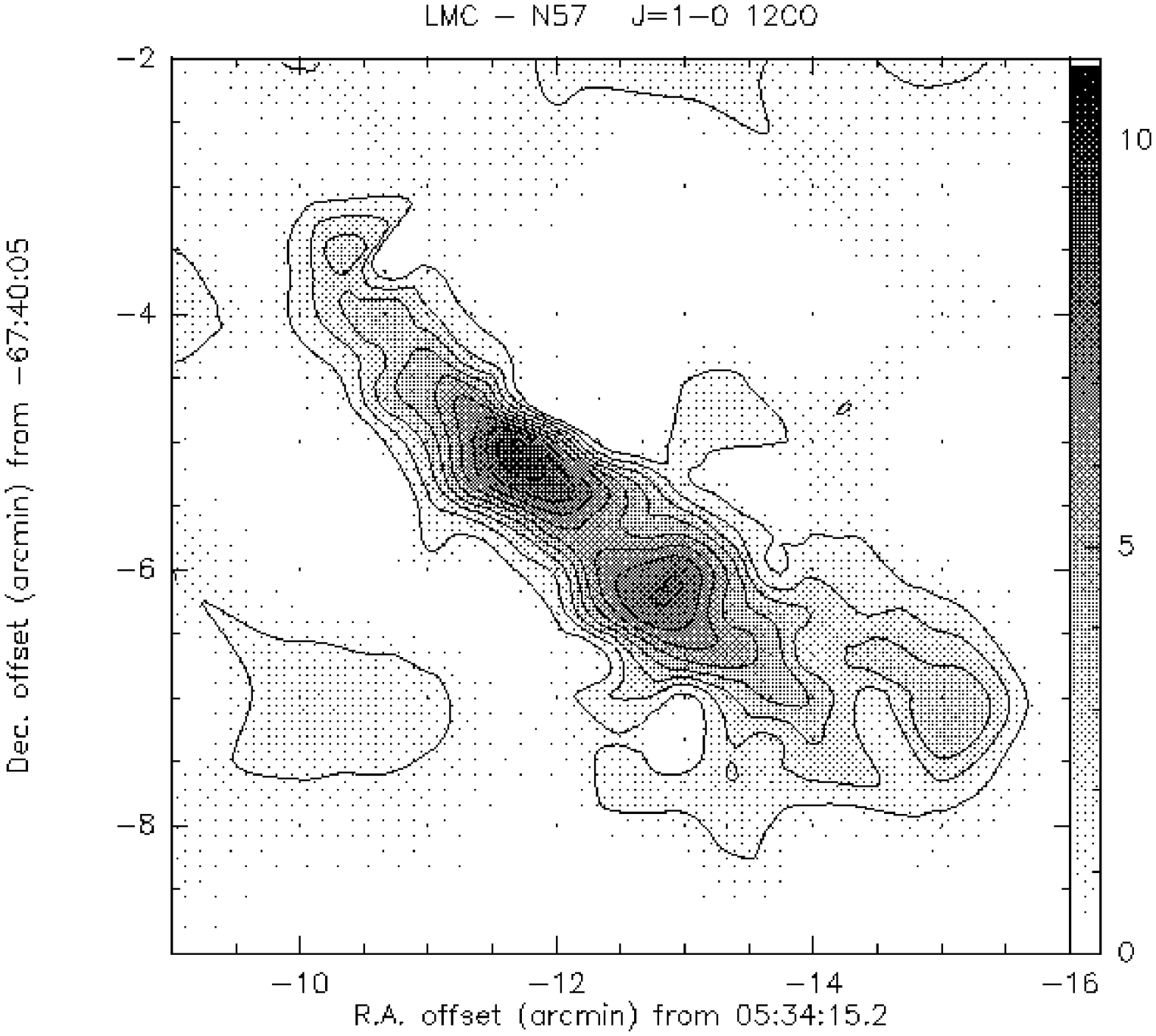}}}
\resizebox{6.cm}{!}{\rotatebox{270}{\includegraphics*{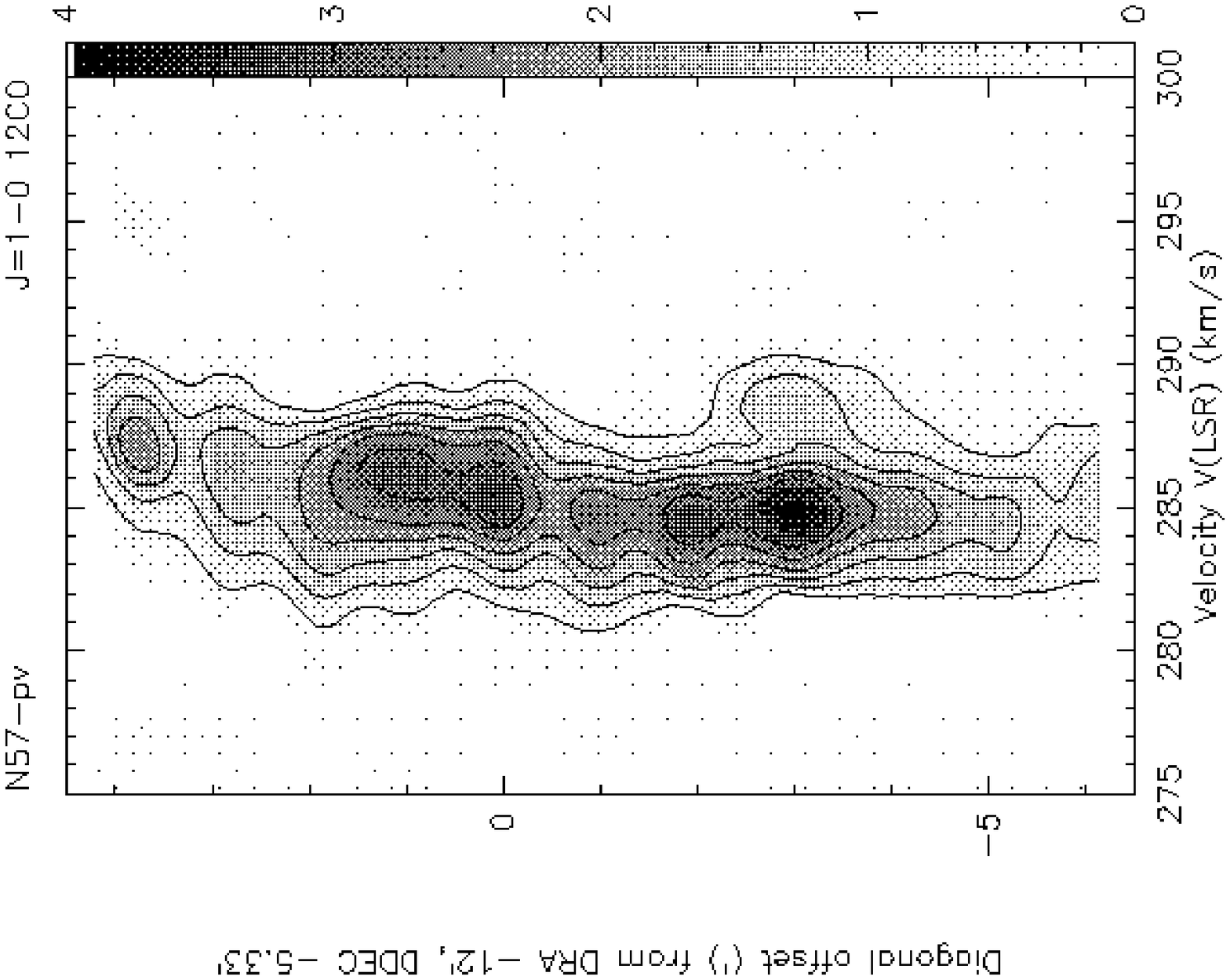}}}
\end{minipage}
\hfill
\begin{minipage}[t]{5.9cm}
\resizebox{6.cm}{!}{\rotatebox{270}{\includegraphics*{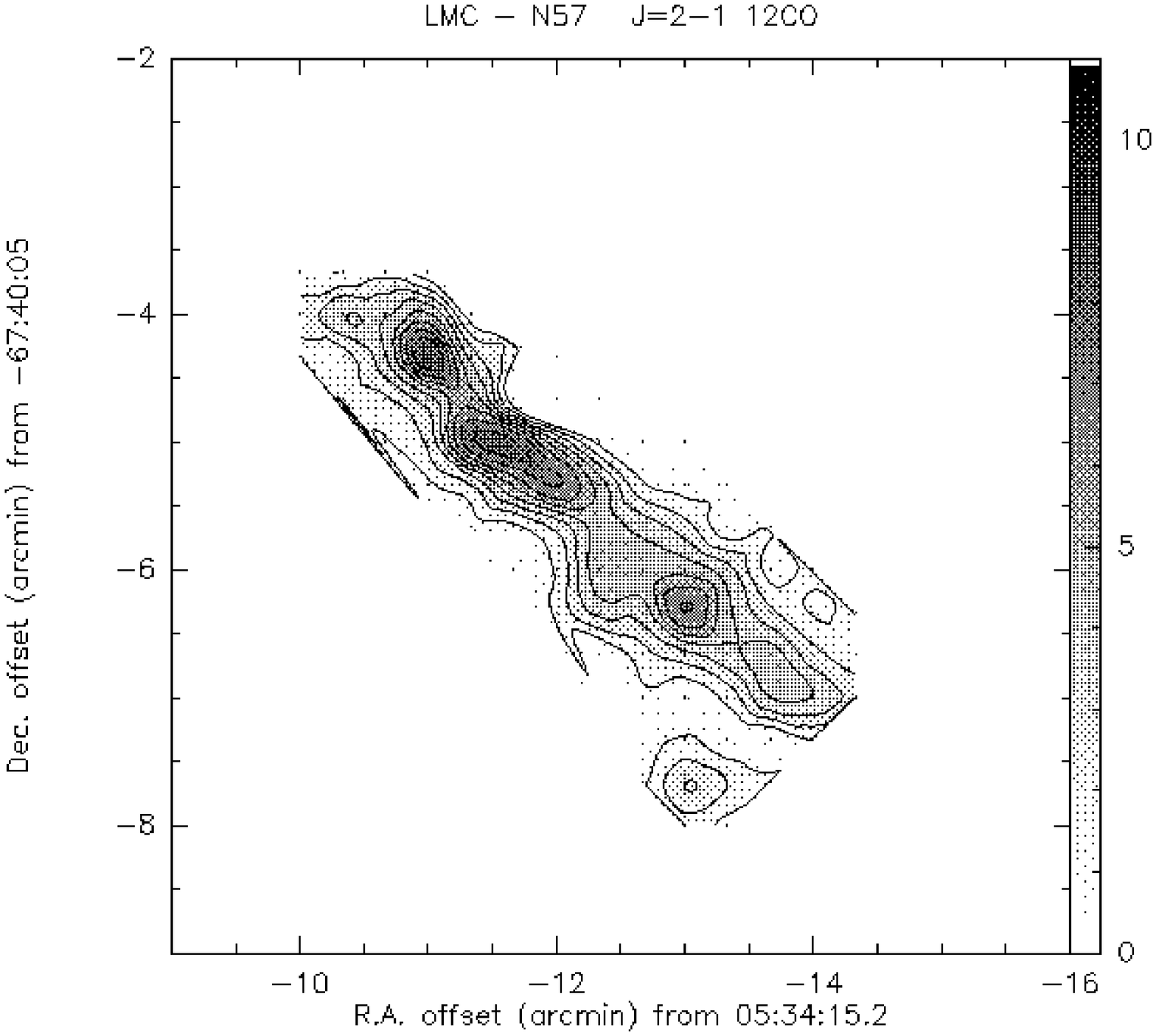}}}
\resizebox{6.cm}{!}{\rotatebox{270}{\includegraphics*{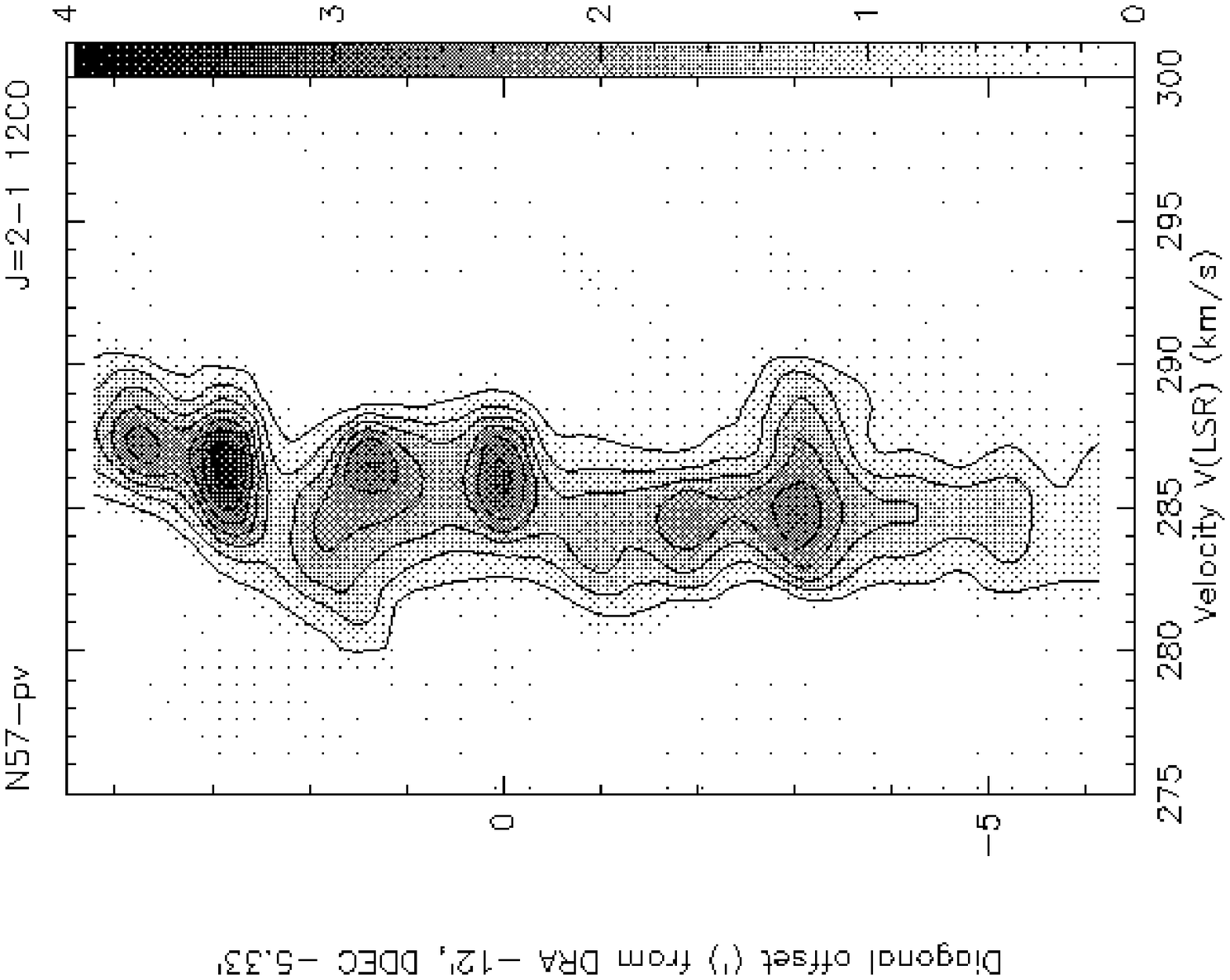}}}
\end{minipage}
\hfill
\begin{minipage}[t]{5.9cm}
\resizebox{6.cm}{!}{\rotatebox{270}{\includegraphics*{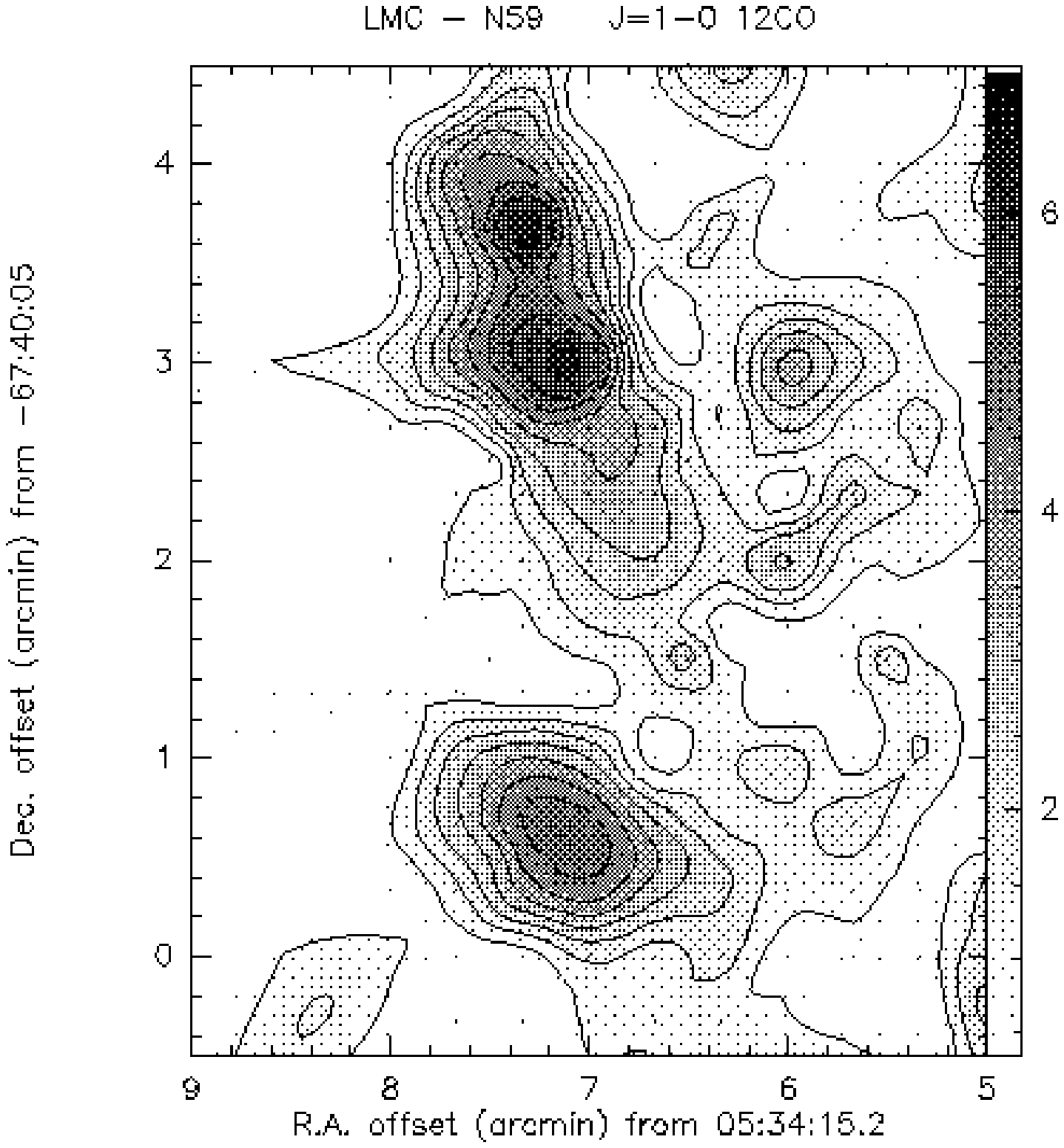}}}
\resizebox{6.cm}{!}{\rotatebox{270}{\includegraphics*{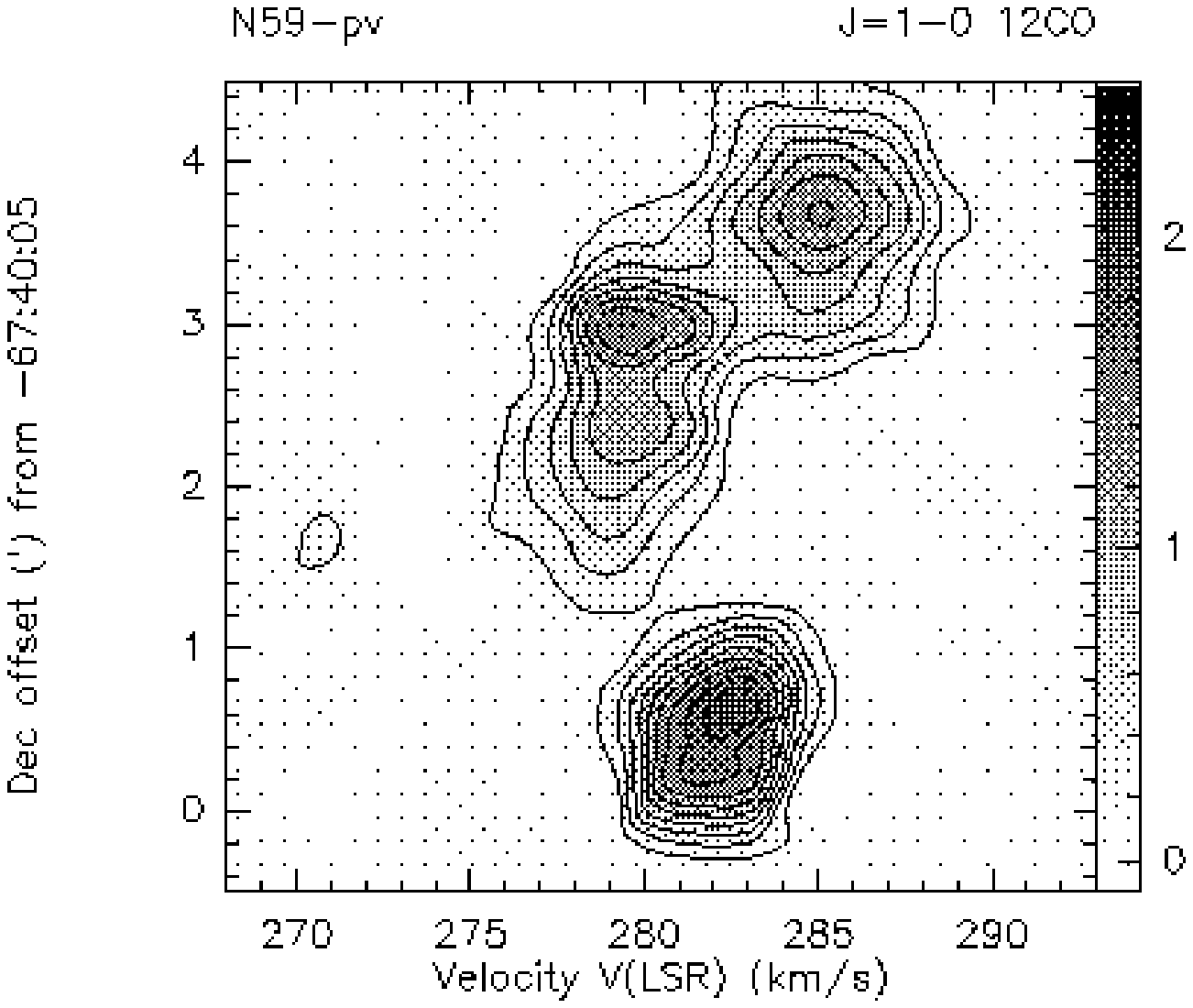}}}
\end{minipage}
\caption[]{Maps of CO clouds associated with LMC HII regions N~57 
and N~59. Linear contours are at multiples of $\int$ $T_{mb}$d$V$ = 
1.4 ($J$=1--0) and 2.0 (J=2--1) $\kkms$ for N~57 and 1.00 $\kkms$ 
for N~59. Position-velocity cut for LMC-N~59 is in declination, 
whereas the cuts for N~57 are diagonal from southwest to northeast. 
Linear contours are at multiples of $T_{mb}$ = 0.70 K ($J$=1--0) and 
1.0 K ($J$=2--1) for N~57 and at 0.30 K for N~59. Gray scales are 
labelled in $T_{A}^{*}$ rather than $T_{mb}$.
}
\label{n57}
\end{figure*} 

\begin{figure*}
\unitlength1cm
\begin{minipage}[t]{17.9cm}
\resizebox{9.cm}{!}{\rotatebox{270}{\includegraphics*{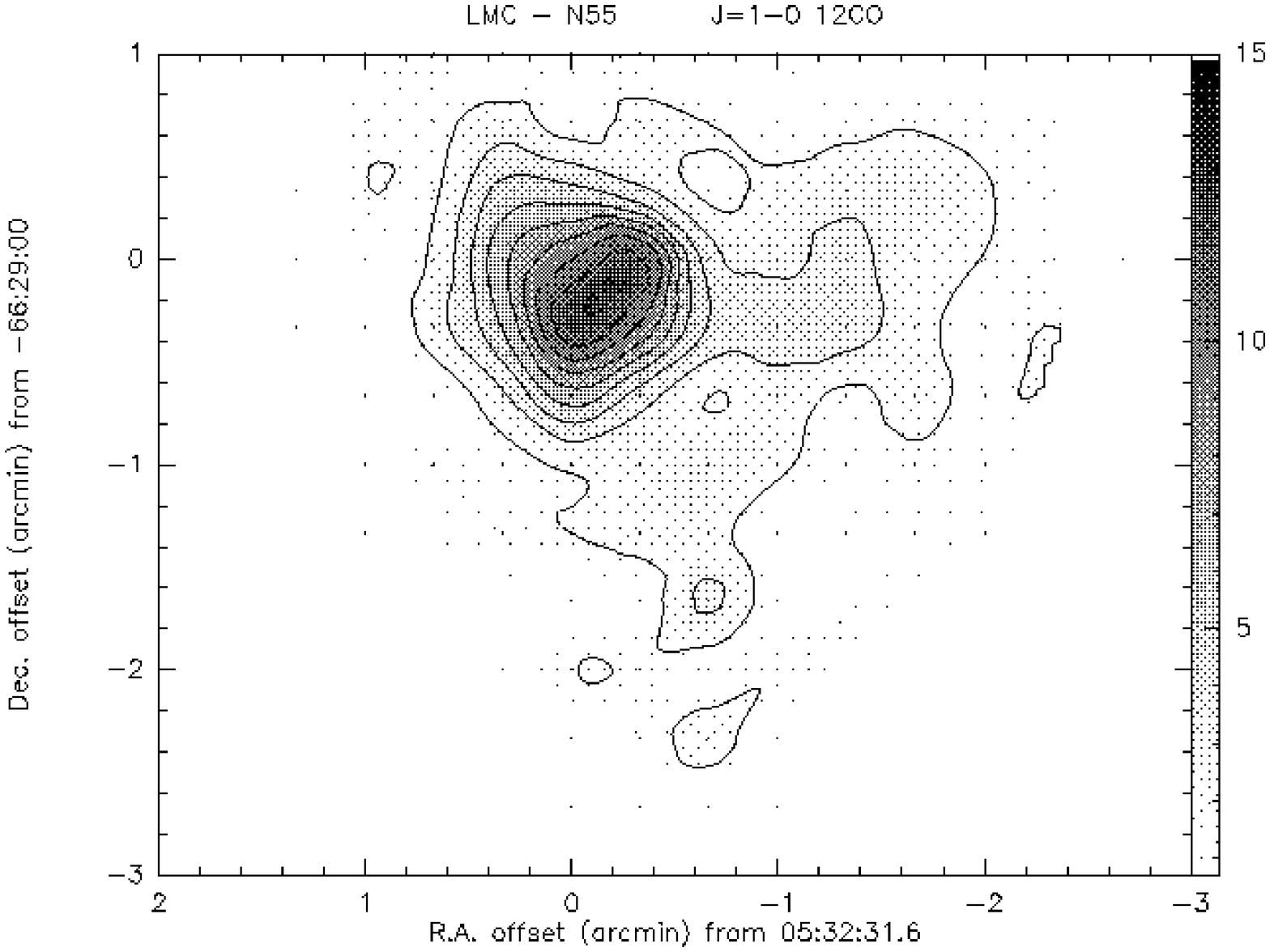}}}
\hfill
\resizebox{9.cm}{!}{\rotatebox{270}{\includegraphics*{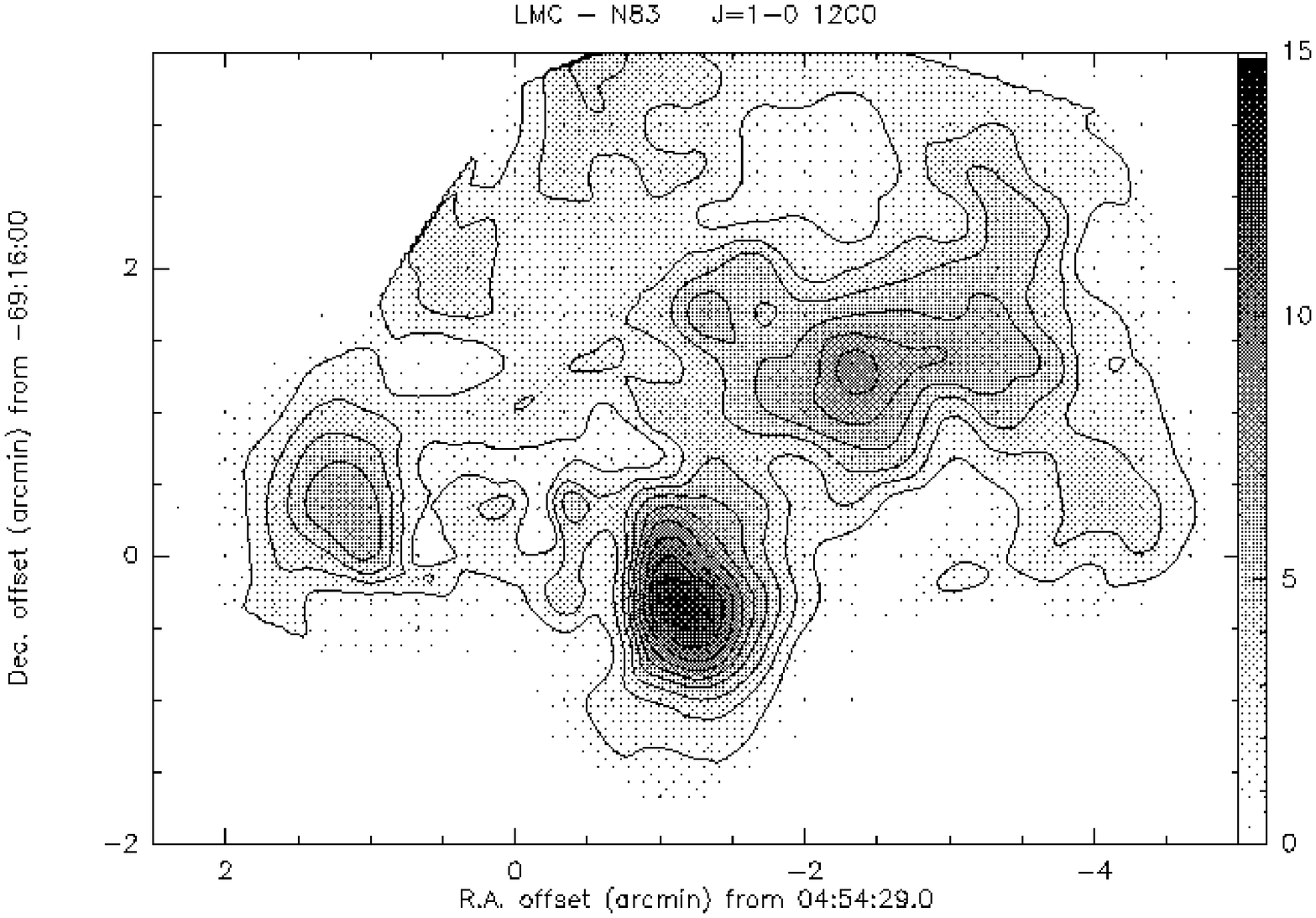}}}
\end{minipage}
\begin{minipage}[t]{17.9cm}
\resizebox{5.9cm}{!}{\rotatebox{270}{\includegraphics*{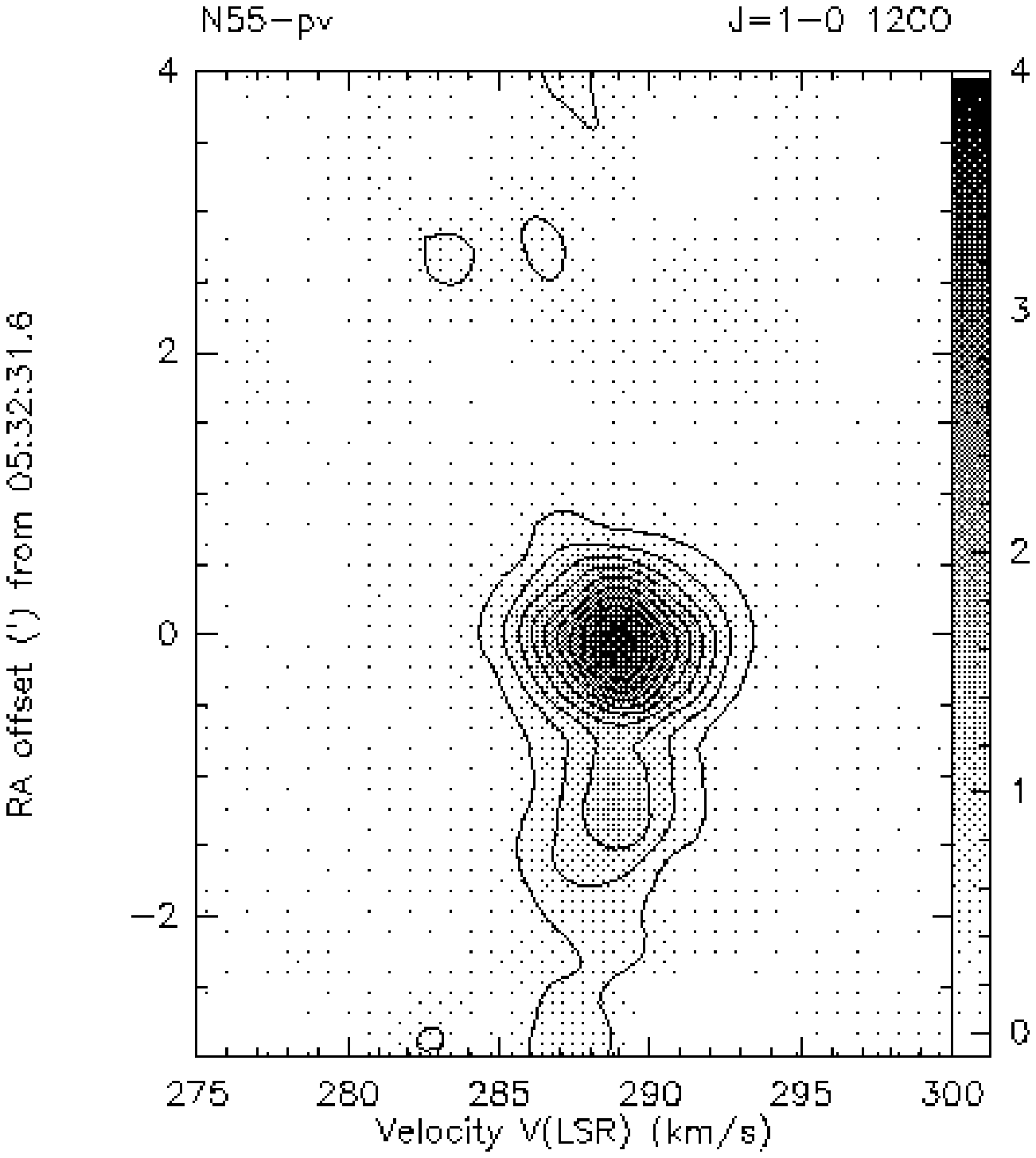}}}
\hfill
\resizebox{5.9cm}{!}{\rotatebox{270}{\includegraphics*{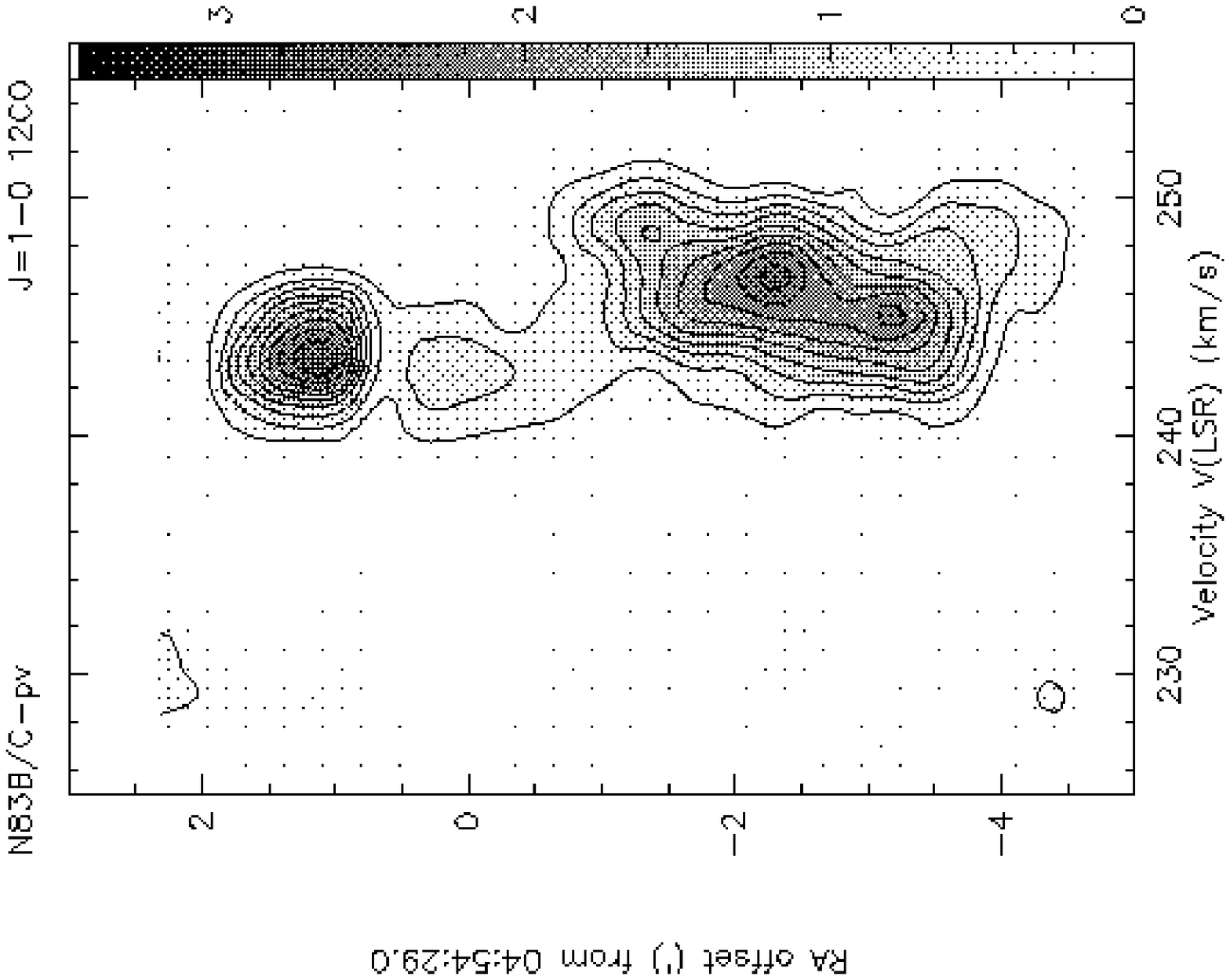}}}
\hfill
\resizebox{5.9cm}{!}{\rotatebox{270}{\includegraphics*{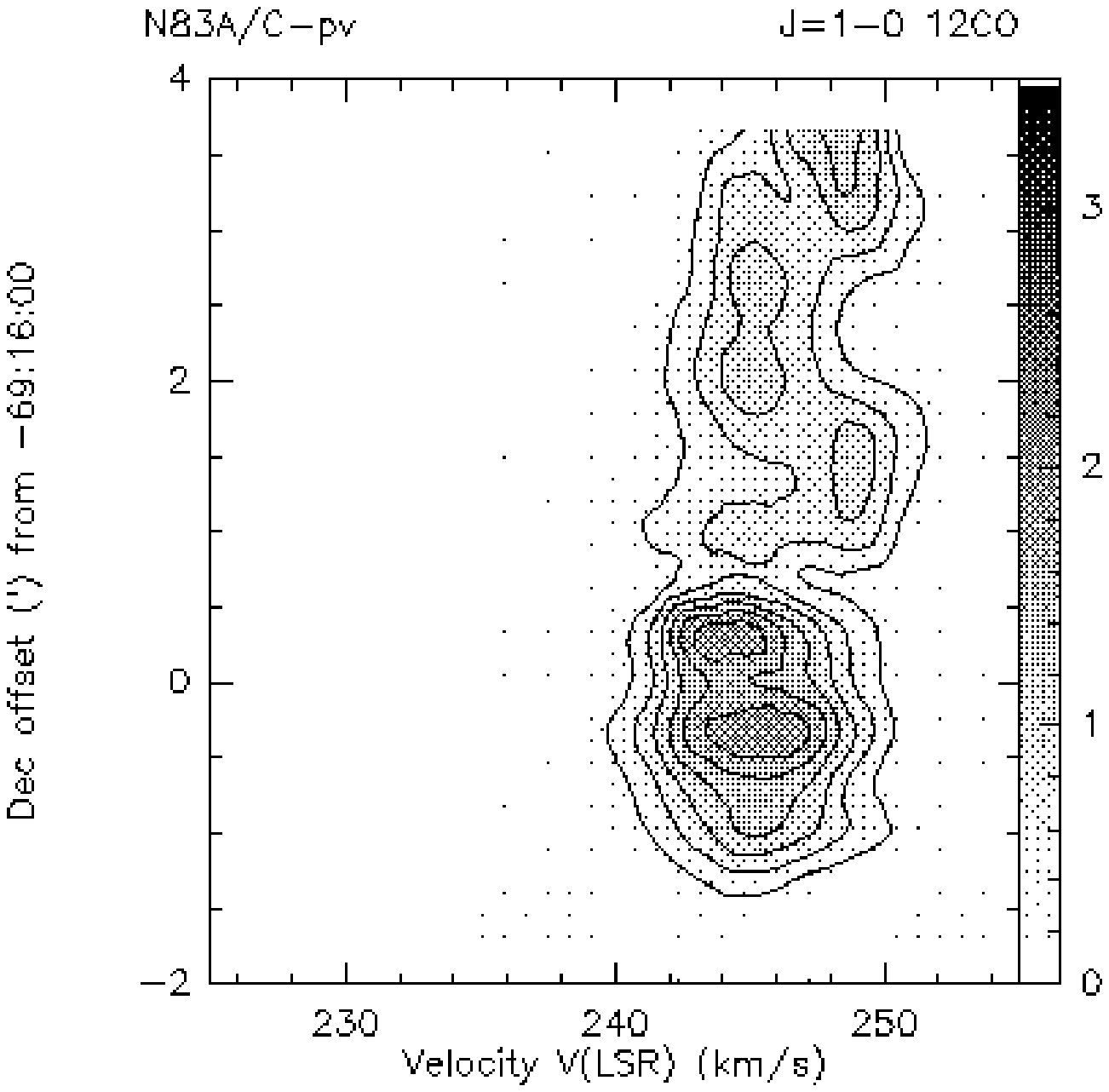}}}
\end{minipage}
\caption[]{Maps of CO clouds associated with LMC HII regions N~55 
and N~83. Linear contours are at multiples of $\int$ $T_{mb}$d$V$ = 
2.0 $\kkms$ for both N~55 and N~83. For both objects position-velocity
cuts in right ascension are shown and for N~83 also a cut in declination. 
The right-ascension cut through N~83 is at a declination offset +1.3 for 
right-ascension offsets below +0.5 and jumps to a declination offset +0.3 
for r.a. offsets above +0.5. Linear contours are at multiples of 
$T_{mb}$ = 0.5 K for N~55 and  at 0.4 and 0.30 K respectively for N~83. 
Gray scales are labelled in $T_{A}^{*}$ rather than $T_{mb}$.
}
\label{n55}
\end{figure*}

In Table~1, we present a list of all Henize (1956) star formation 
regions mapped in the Key Programme. This is a subset of the
HII region sample surveyed in the beginning of the project 
(Israel et al. 1993). Table~1 also includes regions for which the 
results have already been published; it serves as an overall guide 
to Magellanic cloud areas mapped in the SEST Key Programme. 

The observations were made between December 1988 and January 1995
using the SEST 15 m located on La Silla (Chile)\footnote{The 
Swedish-ESO Submillimetre Telescope (SEST) is operated jointly 
by the European Southern Observatory (ESO) and the Swedish Science 
Research Council (NFR).}. Observations in the $J$=1--0 transition
(110 -- 115 GHz) were made with a Schottky receiver, yielding typical 
overall system temperatures $T_{\rm sys}$ = 600 -- 750 K. Observations 
in the $J$=2--1 (220 -- 230 GHz) were made with an SIS mixer, yielding 
typical overall system temperatures $T_{\rm sys}$ = 450 -- 750 K 
depending on weather conditions. On average, we obtained 1$\sigma$
noise figures in a 1 km s$^{-1}$ band of 0.04, 0.10, 0.08 and 0.12 K 
at 110, 115, 220 and 230 GHz respectively.

In both frequency ranges, we used the high resolution acousto-optical 
spectrometers with a channel separation of 43 kHz. The $J$=1--0
observations were made in frequency-switching mode, initially (1988) 
with a throw of 25 MHz, but subsequently with a throw of 15 MHz. 
The $J$=2--1 measurements were made in double beam-switching mode, 
with a throw of 12$'$ to positions verified from the $J$=1--0
$\co$ map to be free of emission. Antenna pointing was checked 
frequently on the SiO maser star R Dor, about 20$^{\circ}$ from the
LMC; r.m.s. pointing was about $3''-4''$. Mapping observations
usually started in the $J$=1--0 $\co$ transition on a grid of $40''$ 
(single-beam) spacing, although in exceptional cases where large
areas were to be surveyed (e.g. Doradus region and N~11 in the LMC), 
double-beam spacings of $80''$ were employed. Where emission was 
detected, we usually refined the grids to a half-beam sampling of 
$20''$. Some of the clouds thus mapped in $J$=1--0 $\co$ were 
observed in $J$=1--0 $\13co$ on the same grid, and with $10''$ 
grid-spacing in the $J$=2--1 transitions.

Because the original observations of LMC cloud N~57 were undersampled
on a grid of 1$'$ spacing, we reobserved in February 2003 that part of 
the map which showed emission from this object. The $J$=1--0 and
$J$=2--1 transitions of $^{12}$CO were observed simultaneously.
The observations were likewise made in frequency-switched mode,
with a throw of 10 MHz, using an autocorrelator for backend. The
resulting new $J$=1--0 observations were combined with the older
ones in a single map.

Unfortunately, frequency-switched spectra suffer from significant 
baseline curvature. In this paper, we have corrected baselines by 
fitting polynomials to them, excluding the range of 
velocities covered by emission and the ranges influenced by negative 
reference features. For each source, the emission velocity range was 
determined by summing all observations, which has the advantage that, 
in principle, it does not select against weak extended emission, at 
least over the same velocity range as occupied by the brighter emission.

The FWHM beams of the SEST are $45''$ and $23''$ respectively at frequencies
of 115 GHz and 230 GHz. Nominal main-beam efficiencies $\eta_{\rm mb}$ at 
these frequencies were 0.72 and 0.57 respectively. For a somewhat more
detailed discussion of the various efficiencies involved, we refer to
Johansson et al. (1998; Paper VII). 

Resulting CO images and position-velocity maps are shown in 
Figs.~\ref{n12},\,\ref{n88},\,\ref{n57} and \ref{n55}; representative
$\co$ and $\13co$ profiles are shown in Fig.~\ref{prof}.
Objects shown include clouds associated with the SMC HII regions
N~12, N~27 and N~88. Profile maps of these clouds, but not
images and position-velocity maps, were earlier presented by 
Rubio et al. (1996).

\section{Results and analysis}

\subsection{Individual cloud properties}

Although all but one of the clouds listed in Table~2 are resolved, 
virtually all of them have dimensions no more than a few times the size 
of the $J$=1--0 $\co$ observing beam (11.2 pc in the LMC and 13.1 pc
in the SMC). The maps thus do not provide much information on the 
actual structure of individual clouds. We determined cloud CO 
luminosities by integrating over the relevant map area. We verified 
that the results were not significantly affected by the precise size 
and velocity limits of the maps. Characteristic cloud dimensions $R'$ were 
determined by counting the number $N$ of map pixels with significant 
emission, and taking $R' = (N/\pi)^{0.5} \Delta S$ where $S$ is the linear
grid spacing. The results were then corrected for finite beamwidth
to yield corrected radii $R$. In Table~2 we list these CO cloud radii 
and luminosities, in addition to the parameters describing the CO 
emission peak. Although it is by no means certain that the clouds 
identified by us are indeed virialized, we have used the data given 
in Table~2 to calculate virial masses following:\\

\noindent
$M_{\rm vir}/{\rm M_{\odot}} = {\rm k}\, R/{\rm pc}\, (\Delta V/{\kms})^{2}$\\

\noindent
where k = 210 for homogeneous spherical clouds and k = 190 for clouds
with density distributions $\propto r^{-1}$  (MacLaren et al. 1988).
In our calculations, we have assumed the former case, although the 
actual uncertainties are in any case much larger than the difference 
between the two values of k. The results are also included in the Table.
As in previous papers, we have searched for correlations between 
source radius $R$, velocity width $\Delta V$, source luminosity $L_{\rm CO}$ 
and virial mass $M_{\rm vir}$. Although the present sample is relatively
small and inhomogeneous, we find that $\Delta V$ and $R$ appear to be
unrelated. There is a marginally significant correlation 
log $L_{\rm CO} \propto$ 2 log $\Delta V$ that appears to be significantly
steeper than the linear correlation found for the N~11 clouds in
Paper IX. However, the more significant correlation between $M_{\rm vir}$
and $L_{\rm CO}$ is within the margin of error identical to the almost
linear correlation found in N~11.

Comparison of the virial masses, corrected for a helium contribution 
of 30$\%$ by mass, with the observed CO luminosity yields, for each
cloud, a mean CO-to-$\h2$ conversion factor $X$, following: \\

\noindent
$X = 1.0 \times 10^{22}\, R\, (\Delta V^{2})\, L_{\rm CO}^{-1}$ \\

\noindent
which is included in the last column of Table 2. 

We find for the {\it discrete CO clouds} a range of $X$ values between 
$2 \times 10^{20}$ and $8 \times 10^{20} \cm2 (\kkms)^{-1}$, with  
effectively identical means $X(SMC)$ = $4.8\pm1.0 \times 10^{20} \cm2 
(\kkms)^{-1}$ and $X(LMC)$ = $4.3\pm0.6 \times 10^{20} \cm2 (
\kkms)^{-1}$, i.e. 2.5 times the `standard' conversion factor in the 
Solar Neighbourhood. Johansson et al. (1998), Garay et al. (2002) and
Israel et al. (2003) obtained very similar results for clouds in 
various LMC complexes (30 Doradus, Complex 37 and N~11 respectively). 

\begin{figure*}
\unitlength1cm
\begin{minipage}[t]{17.9cm}
\resizebox{18.cm}{!}{\rotatebox{270}{\includegraphics*{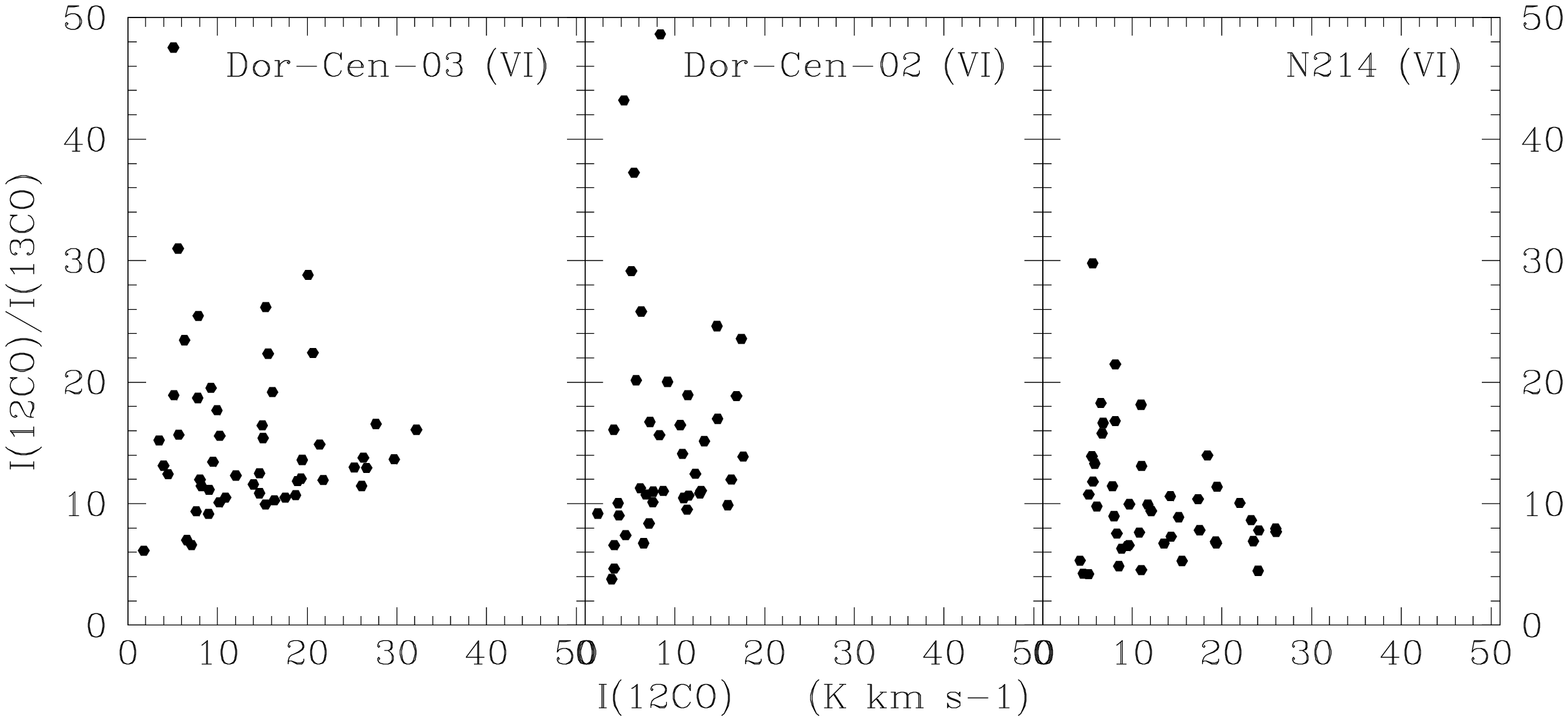}}}
\resizebox{18.cm}{!}{\rotatebox{270}{\includegraphics*{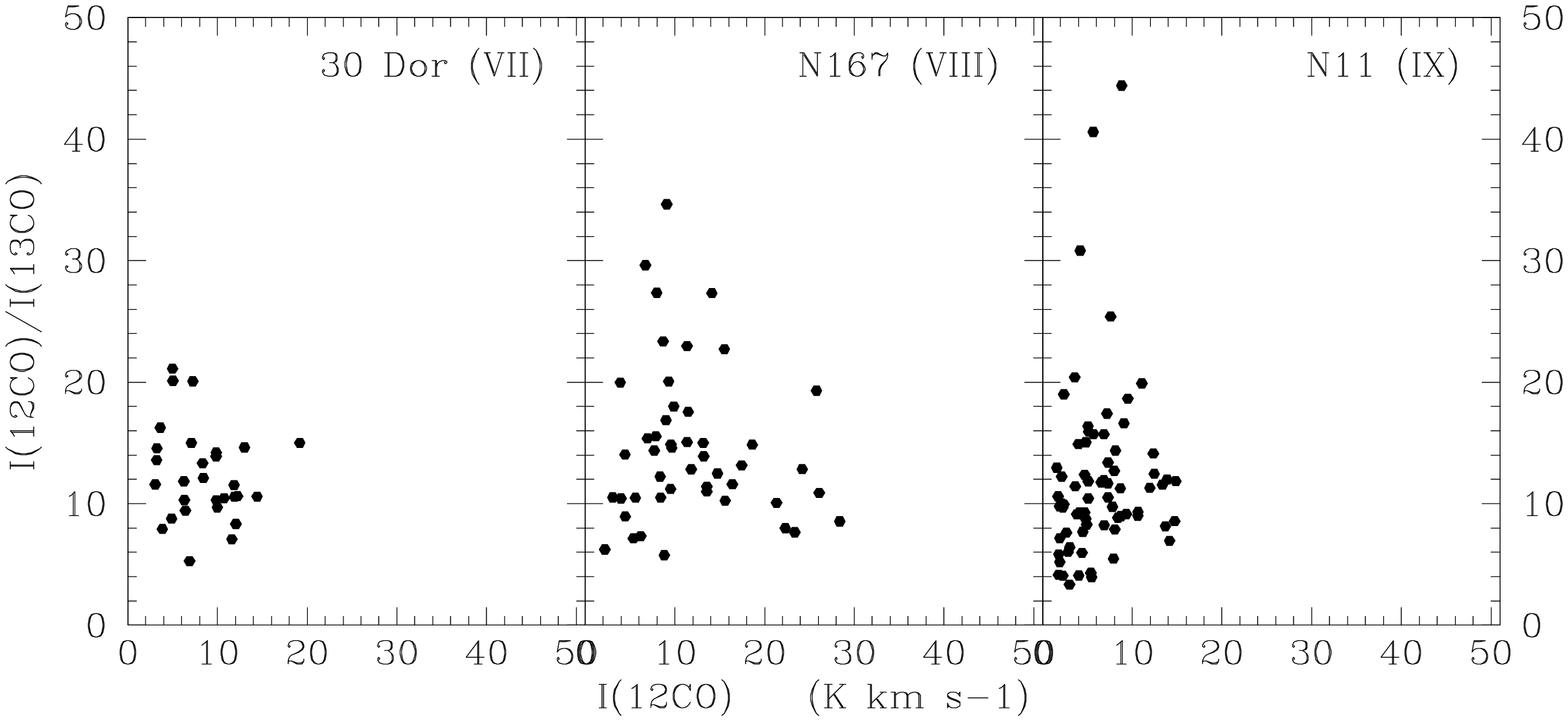}}}
\end{minipage}
\caption[]{The isotopic ratio $I(\co)/I(\13co)$ as a function of velocity 
integrated intensity $I(\co) = \int T_{mb}({\rm CO}){\rm d}V$ for 
$\13co$ pointings in various LMC fields. The observed source is 
identified in each panel, together with a Roman numeral referring
to Key programme paper in which source maps are presented.
}
\label{isorat}
\end{figure*}

\begin{figure*}
\unitlength1cm
\begin{minipage}[t]{17.9cm}
\resizebox{18.cm}{!}{\rotatebox{270}{\includegraphics*{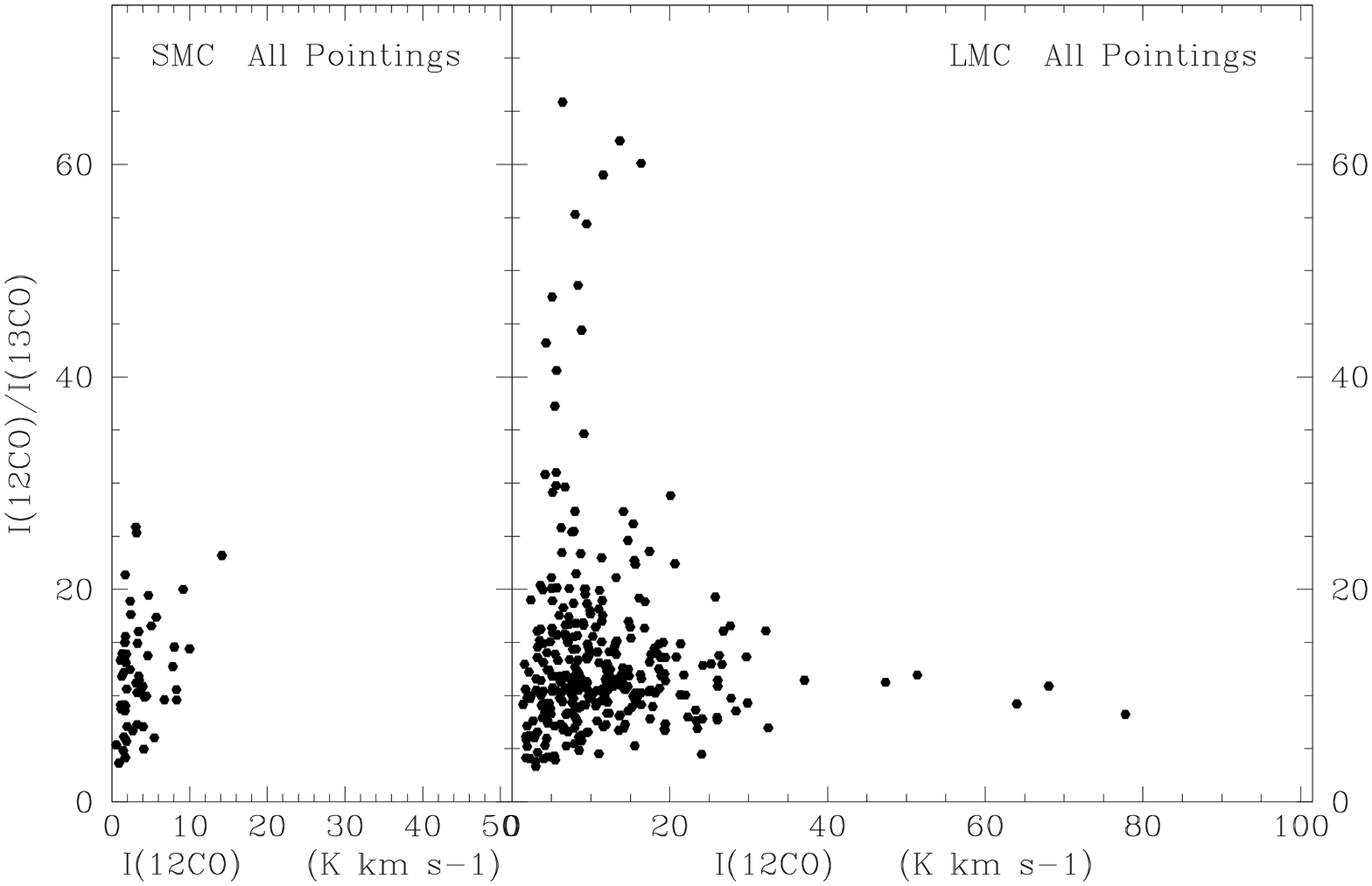}}}
\end{minipage}
\caption[]{The isotopic ratio $I(\co)/I(\13co)$ as a function of velocity 
integrated intensity $I(\co) = \int T_{mb}({\rm CO}){\rm d}V$ for all
$\13co$ pointings in the SMC and the LMC. 
}
\label{mcrat}
\end{figure*}

\subsection{CO Clouds in the SMC}

As the physical characteristics of the molecular clouds associated 
with the Bar HII regions N~12 through N~66 have been discussed in 
previous (Key Programme) papers (see references in Table 1), we refer 
to those papers for further detail. This also applies to the CO 
observations of the cloud associated with the Wing HII region N~88, 
although we note that the previously quoted very high central
$J=1-0 \co/\13co$ isotopic ratio of about 25 was in error and should 
be replaced by half that value as listed in Table~2, rendering N~88
more similar to N~12 and N~27. For the molecular clouds associated 
with the SMC Wing HII region complex N~83/N~84 we refer to a 
forthcoming paper by Bolatto et al. (2003). Here, we will briefly 
comment on the overall characteristics of the CO cloud population of 
the SMC. In Paper I, we concluded that the peak CO emission from 
clouds in the SMC is weak with respect to that from clouds in the 
LMC. This is borne out by the mapping results in Table~2. In fact, 
the brightest CO cloud in the SMC is less conspicuous than 
the brightest subclouds in each of the LMC sources. Yet this object, 
N~27 also known as LIRS~49, is significantly brighter than all 
other sources found in the SMC, including objects not listed here, 
such as N~66 (Rubio et al. 2000) and the various other clouds mapped 
in the southwestern Bar of the SMC (Rubio et al. 1993a; hereafter
Paper II). 

Secondly, it should be noted that the CO clouds have rather small 
dimensions. This is not only shown by the velocity-integrated 
intensity maps, but it is also quite obvious from the various 
position-velocity maps that invariably show very limited extents 
for bright emission regions. Bright regions extending over more 
than an arcminute (18 pc) always show substructure suggesting  
that the observed source is a grouping of smaller 
individual clouds (see also Figs.~1 through 4 in Paper II). 
The small CO clouds associated with the Orion-sized HII 
regions N~12 and N~27 in the southwest main Bar of the SMC and 
N~88 in the SMC Wing, are comparatively simple and not part of 
a larger complex. This is in contrast to the clouds associated
with N~66 (main Bar) and N~83/N~84 (Wing). These occur in 
complexes up to 4 arcmin (70 pc) in size as do the clouds in 
the southwest Bar (SMC-B1 and SMC-B2 in Paper II). 
With the exception of the N~27 cloud, all CO (sub)clouds mapped 
in the SMC, including the N~66 complex, are {\it significantly 
smaller in size} than the associated HII regions or HII region 
complexes. This situation is unlike that found in the Milky Way, 
where CO cloud complexes are frequently much larger than the 
associated HII regions. Moreover, as Figs.~1 and 3 in Paper II
show, the HII regions, although larger than the nearest CO
clouds, are usually centered at the edge of these clouds 
(N~15, N~16, DEM~34 and DEM~35; N~25), or occur at a hole in
the CO distribution (N~13, N~22). This is true also for the
HII regions whose CO maps are presented here, N~12, N~27 and 
N~88, which are centered at the western, eastern and southwestern
edges of their respective CO clouds. Of particular interest in 
this respect are N~27 and N~83 where the CO emission appears to
occur predominantly in a ridge adjacent to the HII region.

Although the extended emission from N~88 is centered at about
$\Delta \alpha = 0, \Delta \delta = -0.33$ in the map shown
in Fig. ~\ref{n88}, the dust-rich, high-excitation compact 
component N~88A (see Heydari-Malayeri et al. 1999) occurs close 
to the map center, where the CO emission exhibits an extension to 
the northwest. Shocked molecular hydrogen was detected in this
nebula, which appears to be partially embedded in the molecular
cloud (Israel $\&$ Koornneef 1991).

Both the limited size of the clouds 
and their weak CO emission causes the resulting CO luminosities 
to be rather modest. Most of the clouds identified in the SMC 
Bar region have luminosities $L_{\rm CO} = 1000 - 2000 \kkms$ 
pc$^{2}$ (Paper II); the N~88 cloud in the Wing has a very low 
$L_{\rm CO} \approx 500 \kkms$ pc$^{2}$. Cloud complexes such as 
N~66 (Rubio et al. 2000) in the Bar and N~83/N~84 in the Wing 
typically have equally modest luminosities $L_{\rm CO} \approx 
6000 \kkms$ pc$^{2}$. Relatively small HII regions in the Bar 
have the brightest CO clouds: N~12 and N~22 (= SMC-B2 no.3, 
Paper II) both have $L_{\rm CO} \approx 4000 \kkms$ pc$^{2}$, 
and N~27 has $L_{\rm CO} \approx 8500 \kkms$ pc$^{2}$.
The identified clouds represent a significant fraction
of the CO present in the SMC, as is clear from a comparison
with the 2$^{'}$.6 (50 pc) resolution maps published by Mizuno 
et al. (2001). At this lower resolution, their survey covers 
a larger surface area in otherwise the same parts of the SMC 
that we have mapped. Nothwithstanding a much more
extensive coverage, their Fig. 1 clearly shows that there is 
little CO emission in the southwest Bar in 
addition to the sources SMC-B1, SMC-B2, LIRS 36 (N~12) and 
LIRS 49 (N~27) mapped by us. Likewise, in the northern Bar 
there is not much emission apart from that associated with 
N~66 and N~76. From their more complete map of the N~83/N~84 
region in the Wing, they obtain $L_{\rm CO} \approx 13 000 
\kkms$ pc$^{2}$, i.e. only twice the value we find in a few 
discrete clouds. Similarly, the total CO luminosity detected 
in their survey is also about twice the sum of the luminosities 
of the individual sources detected by us (Paper II, This Paper).
 
\subsection{CO clouds in the LMC}

Again, most of the Key Programme sources observed in the LMC
and listed in Table~1 have already been discussed in some 
detail in the references given in that Table. The exceptions 
are the sources associated with the HII region complex N~83 
in the center-west of the LMC, and the HII regions N~55, N~57 
and N~59, all associated with supergiant shell SGS-4 (Meaburn 
1980) in the northeast of the LMC. Located between SGS-4 and
SGS-5 are the CO cloud counterparts of HII regions N~48 and 
N~49. These have also been mapped with the SEST by Yamaguchi
et al. (2001b), but not as part of the Key Programme.

\subsubsection{N~55, N~57 and N~59} 

N~55, N~57 and N~59 are all large (respectively 6$'$, 10$'$ 
and 8$'$) HII region complexes associated with supergiant shell 
SGS~4 (Meaburn 1980; see optical image by Braun et al. 1997 or
CO map by Yamaguchi et al. 2001b). The shell and the dominant
stellar population associated with it are 10--30 million years 
old (see references reviewed by Olsen et al. 2001).  The HII 
region complex N~57 is excited by the OB association LH~76 (Lucke 
$\&$ Hodge 1970). N~57 and N~59 are at the southeastern edge of
SGS~4, in a region of the LMC also known as 
Shapley Constellation III (McKibben Nail $\&$ Shapley 1953). In 
contrast, N~55 is seen projected inside the shell. A direct 
physical association of N~57 and N~59 with the supershell is 
suggested not only by the fact that they occur at the shell 
edge, but also by the fact that the CO clouds (Fig.~\ref{n57}) 
form elongated structures almost exactly along this edge. This
configuration is particularly striking for N~57. The elongated 
CO complex contains over half a dozen virtually unresolved
(i.e. radii of only a few parsec) compact components, which
are particularly well-distinguished in the position-velocity
maps shownnin Fig.\ref{n57}. The map containing N~59 likewise
shows at least four distinct CO clouds, the northernmost of 
which is close to the center of the much larger ($2'-3'$) HII 
region N~59A. It is remarkable that both in the case of N~57 and 
of N~59, the CO emission apears to be `sandwiched' between the 
shell and the brightest HII regions. With the present observations, 
it is difficult to retrieve detailed information on the 
physical condition of the clouds.

N~55 is a fairly isolated HII region complex inside the shell. 
It is excited by OB association LH~72. The stellar population, 
and the neutral hydrogen surrounding the object were the 
subject of a detailed study by Olsen et al. (2001). The extent 
of the ionized gas once again greatly exceeds that of the CO 
shown in Fig.~\ref{n55}. Comparison with Figs.~14 and 15 by 
Olsen et al. (2001) suggests that the CO is found at velocities 
devoid of HI emission and predominantly between the ionized 
region and the southernmost peak in the extended HI distribution. 
The appearance of the HI, CO and H$\alpha$ distributions lends 
support to the surmise by Yamaguchi et al. (2001b) that the N~55 
complex has been shaped by the passage of the SGS~4 shell.

Comparison with the CO cloud luminosities resulting from the 
$2.^{'}6$ beam survey by Mizuno et al. (2001) show that 
essentially all CO for at least N~55 and N~57 has been detected
by us. N~59 does not occur in their catalog, although it
is clearly visible in the more sensitive CO map by Yamaguchi 
et al. (2001b).

\subsubsection{N~83}

N~83 is a large HII region complex of diameter $6'\times5'$
(95$\times$80 pc) extending beyond the boundaries of the map in
Fig.~\ref{n55}. It is located at the edge of Meaburn's (1980) 
supergiant shell SGS~6 (see Fig.~1 by Yamaguchi et al. 2001a).
N~83 contains a number of individual bright HII
regions. N~83A is associated with the bright CO cloud in the
lower center of the map; the CO cloud is smaller than the
1.5$'$ (23 pc) diameter of the very bright HII region. The
likewise very bright N~83B is centered at the western edge of
the easternmost CO cloud in Fig.~\ref{n55}; with a diameter
of 0.6$'$ (9 pc) it is somewhat smaller than its associated 
CO cloud. N~83C and N~83D are relatively compact
HII regions (sizes of 0.3$'$ -- 0.4$'$) located at minima in
the CO map inbetween the brighter regions. Although we have
mapped clear cloud edges in the south and west, we cannot 
exclude the possibility that CO emission extends to the 
northeast. However, the CO luminosity detected in our map,
$L_{\rm CO} = 1.95 \times 10^{4} \kkms$pc$^{2}$, is about 
90$\%$ of the cloud luminosity determined by Mizuno et al. 
(2001) from their large-beam ($2^{'}.6$) CO survey of the 
LMC, suggesting that we have not missed much.

\subsection{Isotopic ratios}

\begin{figure}
\unitlength1cm
\begin{minipage}[t]{8.1cm}
\resizebox{8.cm}{!}{\rotatebox{270}{\includegraphics*{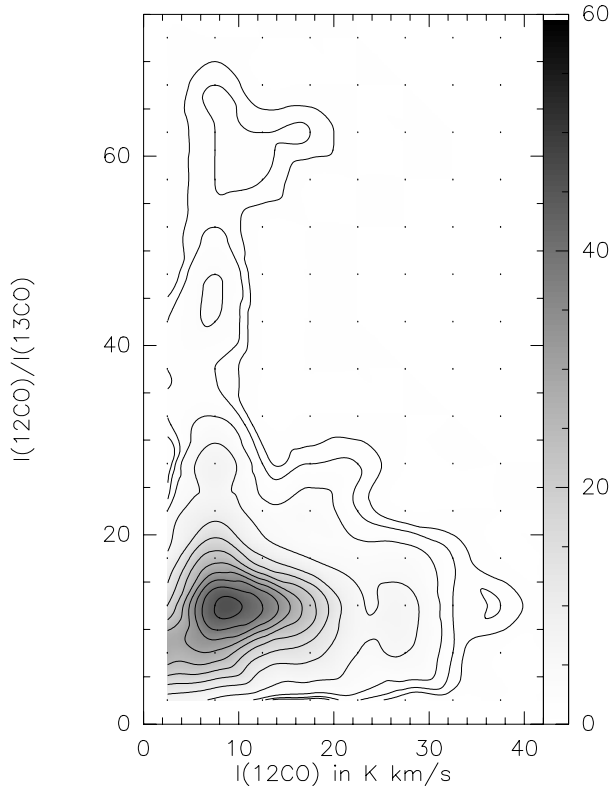}}}
\end{minipage}
\caption[]{The distribution of points in the densely populated
lower left part of Fig~\ref{mcrat} is shown here as an iso-density
contour map. 
}
\label{conrat}
\end{figure}

\begin{figure}
\unitlength1cm
\begin{minipage}[t]{8.1cm}
\resizebox{8.cm}{!}{\rotatebox{270}{\includegraphics*{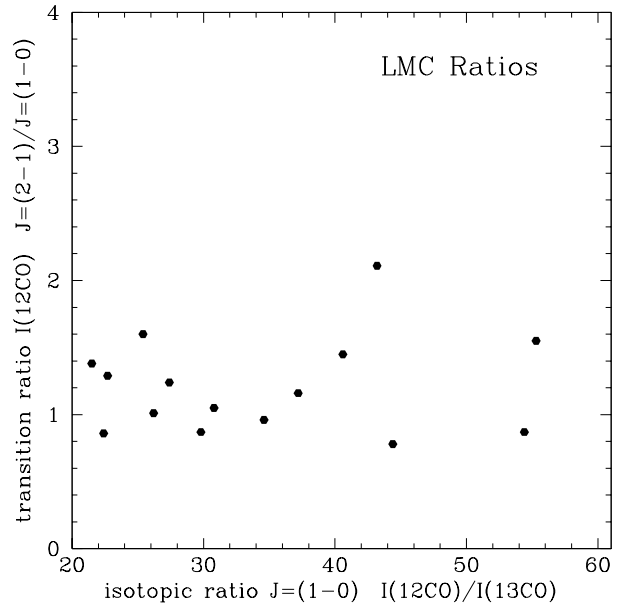}}}
\end{minipage}
\caption[]{Plot of $J=1-0/J=2-1 \co$ transition ratios as a function
of (the higher) $J=1-0 \co/\13co$ isotopic ratios. The transition 
ratios on average slightly exceed unity. 
}
\label{highrat}
\end{figure}

\begin{figure}
\unitlength1cm
\begin{minipage}[t]{8.1cm}
\resizebox{8.cm}{!}{\rotatebox{0}{\includegraphics*{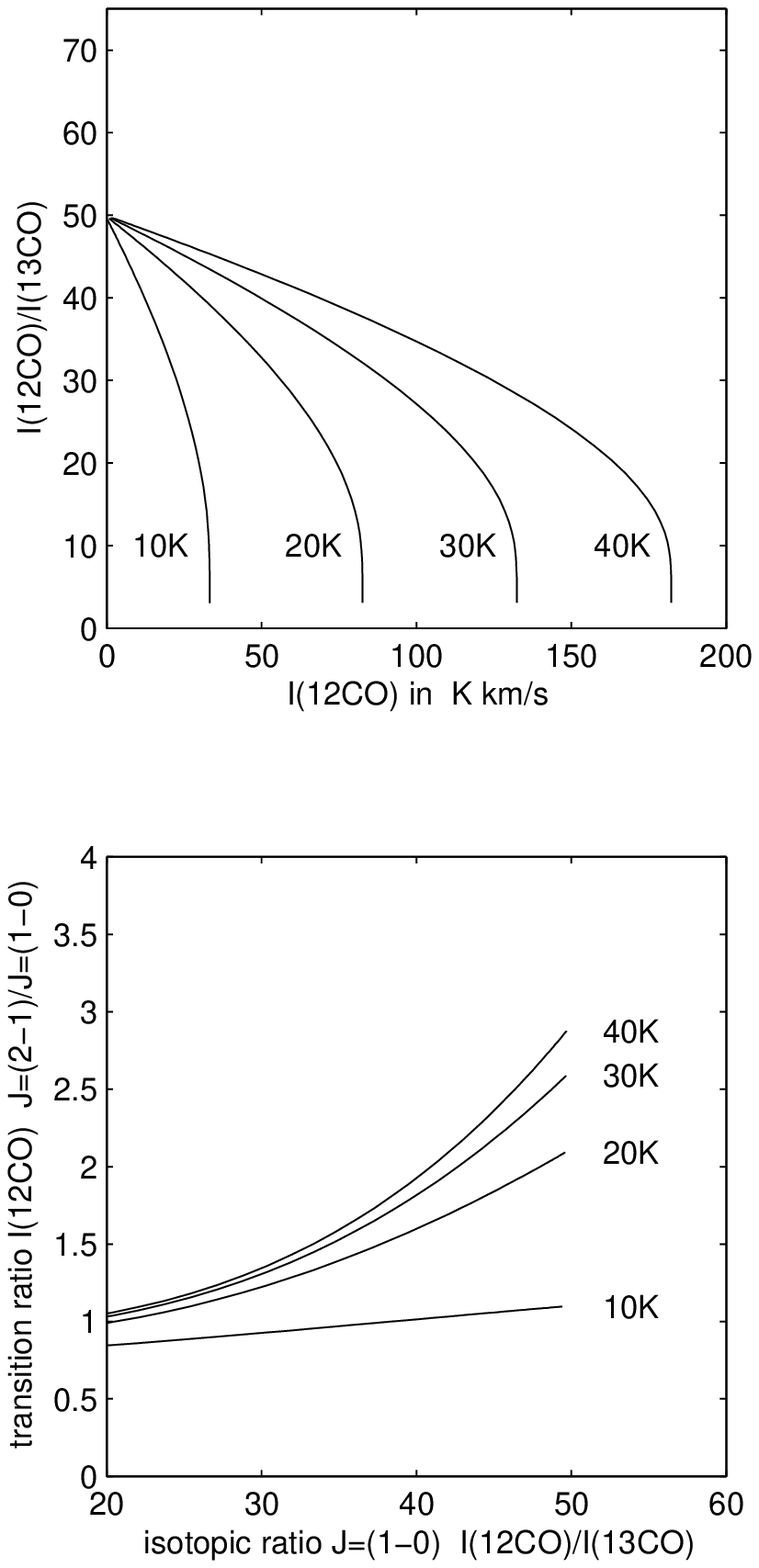}}}
\end{minipage}
\caption[]{Plots showing $\co$ and $\13co$ line ratios as a 
result of LTE modelling of various conditions representing the
Magellanic Clouds environment.
}
\label{lebfig}
\end{figure}

In addition to the $J$=1-0 {\it isotopic ratios} $R_{13} = 
I(\co)/I(\13co)$ corresponding to the peak emission of the 
sources listed in Table~2, we have collected similar ratios 
for all pointings where $J=1-0 \13co$ velocity-integrated 
emission was measured with uncertainties less than 20$\%$. 
In Fig.~\ref{isorat} we display the results for molecular
clouds in the relatively quiescent area south of 30 Doradus 
and N~159 (Kutner et al. 1997, Paper VI), the molecular cloud
associated with the northern ionization front of 30 Doradus
itself (Johansson et al. 1998; Paper VII), the brightest
clouds associated with N~167, to the east of 30 Doradus
(Garay et al. 2002; Paper VIII) and molecular clouds Nos.
8, 10 and 15 forming part of the ring in the N~11 complex 
(Israel et al. 2003; Paper IX). 

A comparison of the fields in Fig~\ref{isorat} illustrates 
interesting differences. For instance, only a limited range 
of CO intensities occurs in N~11 (as pointed out in Paper IX)
but the range of isotopic ratios in this object is large.
In contrast, very high isotopic ratios are absent in the 30 
Doradus cloud. In Fig~\ref{mcrat} we show all available
pointings in the SMC and the LMC.  With a few exceptions,
the SMC measurements are characterized by relatively low 
intensities $I(\co) < 10 \kkms$. Isotopic ratios range from 
about 5 to 25. The LMC results show a richer pattern. 
This diagram contains a number of pointings on molecular
clouds in intensely star-forming complexes such as N~159, 
N~44 and N~214, characterized by CO intensities $I(\co) > 
30 \kkms$. These all have very similar isotopic ratios $
R_{13} \approx 10$, which is about a factor of two higher 
than the isotopic ratios of Galactic molecular cloud centers.  
Throughout the Magellanic Clouds, {\it transitional ratios} 
$r_{21} = (J=2-1)\co/(J=1-0)\co$ are found to be close to 
unity (Papers V, VII, VIII, IX; Rubio et al. 2000). This
all but rules out very low intrinsic $\co$ optical depths. 
The observed $\co$ emission must at least be saturated. 
This implies {\it either (i) lower CO abundances, or 
(ii) a lesser filling of the beam by $\13co$ than by 
$\co$, or (iii) an intrinsically lower $\13co/\co$ 
abundance ratio.} However, the last possibility (iii) does 
not seem to applicable according to estimates by Johansson 
et al. (1994).

For pointings in less bright directions, with $I(\co) 
< 30 \kkms$, the range of isotopic ratios rapidly increases 
from low values of 4 to high values of 70.  
As we only included $\13co$ measurements with reasonable to
good signal-to-noise ratios, this is an intrinsic increase 
in range, {\it not} caused by higher noise levels. In order
to better study the behaviour of ratios and intensities in 
the most densely populated part of Fig~\ref{mcrat}, we
have produced a plot with iso-density contours of that part,
shown in Fig~\ref{conrat}. A full analysis of the CO line 
emission in terms of source structure unfortunately requires 
more transitions than we have observed. However, from
Fig~\ref{conrat} it is obvious that the great majority of 
registered pointings show an isotopic ratio between 10 and
15 that appears to drop slowly as $\co$ intensities decrease. 
This behaviour can be understood as due to relatively cold 
molecular gas having lower brightness
temperatures as well as higher $\co$ {\it and} $\13co$ optical 
depths. At intensities $I(\co) < 20 \kkms$ we find, in addition, 
a relatively small but significant number of pointings that
combine low CO intensities with high isotopic 
ratios. Almost all of these are in the direction of the 
molecular cloud edges; the LMC and SMC molecular clouds
mapped in $\13co$ are smaller than the $\co$ extent. 
The high isotopical ratios are caused either by {\it low 
optical depths in both $\13co$ and $\co$} or by {\it 
$\13co$ filling less of the observed surface area than 
$\co$}. 

In Fig~\ref{highrat} we show transition ratios
$r_{21}$ as a function of the isotopic ratio $R_{13}$. 
We have as much as possible convolved $J$=2--1 $\co$
observations to the twice larger $J$=1--0 beam. This
was not always fully possible. As a consequence, about
half of the transition ratios $r_{21}$ in Fig~\ref{highrat} 
are upper limits although we believe that they are
usually quite close to the actual value. The average 
$J$=2--1/$J$=1--0 transition ratio is about 1.2. 

Assuming a $\13co/\co$ abundance ratio of about 50
(Johansson et al. 1994) and CO rotation temperatures
less than 30 K, Fig~\ref{highrat} does not distinguish
between very high (as in the Galaxy) and moderately high
($\tau \leq 5$) optical depths in the $J=1-0 \co$ line.
However, Fig.~\ref{lebfig} shows that the outer envelope of 
the distribution shown in Fig.~\ref{conrat} is well fitted 
by a line of constant $T_{\rm rot}$ ($\approx 12$ K) 
defined by varying optical depth. This applies 
particularly to the region below an isotopic ratio of 
about 40, under the assumption of LTE conditions, an 
intrinsic isotopic abundance ratio of 50, and full
beam-filling. For beam-filling factors less than unity, 
the best-fit $T_{\rm rot}$ increases. For a factor of 0.5, 
for instance, we find $T_{\rm rot} \approx 20$ K. The
extension towards even higher isotopic ratios,
discernible in Fig~\ref{conrat}, suggests a difference 
in the $\13co$ and $\co$ beam-filling.
This would be the natural consequence of vigorous CO 
photodissociation expected to occur in the UV-rich and 
metal-poor environment of star formation regions in the 
Magellanic Clouds. Both $\co$ and $\13co$ would be 
affected by the lack of shielding against erosion by 
energetic photons, producing weaker emission in a given
beam. However, the lower abundance $\13co$ isotope
would have much less self-shielding by its significantly 
lower column-density and therefore would suffer much
more dissociation. Consequently, one expects
isotopic ratios to increase with decreasing $\co$
intensities. The lack of high ratios at the very lowest
CO intensities in Figs~\ref{mcrat} and \ref{conrat} is
a selection effect: the very low $\13co$ intensities
implied by those ratios have been excluded by our
requirement of an acceptable signal-to-noise ratio.

In summary, our data indicate that two of the three 
explanations suggested are actually at work: (i) lower 
CO abundances in the Magellanic Clouds with respect to 
Galactic clouds and (ii) different filling factors for 
$\13co$ and $\co$, at least in low-density regions.
The first point naturally explains the higher
$\co$/$\13co$ intensity ratios observed in the Magellanic
Clouds, even though the intrinsic isotopic ratio 
seems similar to that in the Galaxy.

\section{Conclusions}

\begin{enumerate}
\item We present high-resolution (40$''$ corresponding to
about 11 pc) maps of molecular cloud CO emission associated
with 5 star-forming regions in the SMC (N~12, N~27, N~83, 
N~84 and N~88) and 4 in the LMC (N~83, N~55, N~57 and N~59),
in addition to the 7 respectively 8 already published
in previous papers in this series (see Table 1).
\item Most of the clouds detected are resolved, but not
much larger than the observing beam. Almost all the
CO clouds mapped are often much smaller than the
extent of the associated ionized gas (HII region).
\item If we assume these
clouds to be virialized, the resulting masses define
a poor but significant linear correlation with J=1-0
$\co$ luminosity. Molecular cloud masses thus derived
lie typically between 1000 and 5000 $\Msun$, although
much lower and much higher masses both occur.
\item Under the assumption of virialization, the small
discrete clouds mapped have CO-to$\h2$ conversion factors
 $X$ that are only 2--3 times the conversion factor in 
the Solar Neighbourhood. The widespread lack of diffuse
CO emission, among others, suggests that most clouds
are part of photon-dominated regions (PDR's) so that
total molecular masses, incompletely traced by CO, may 
be higher.
\item We have also collected all detections of the
$J=1-0 \13co$ transition in the Key Programme, including 
those from sources published in previous papers. 
Isotopic ratios $I(\co)/I(\13co)$ of the majority of
these detections cluster around a value of 10. We believe 
this to reflect substantially lower CO abundances in the 
Magellanic Clouds, commensurate with the low-metallicity 
strong-radiation ambient environment. 
\item At low $\co$ intensities we also find isotopic
ratios both lower and higher than the above value of 10.
We attribute the former to relatively cool and dense 
molecular gas, and the latter to cloud edges particularly
strongly affected by CO photo-destruction.
\end{enumerate}

\acknowledgements

It is a pleasure to thank the operating personnel of the SEST 
for their support, and Alberto Bolatto for valuable assistance 
in the reduction stage. M.R. wishes to acknowledge support from
FONDECYT through grants No 1990881 and No 7990042.


\begin{thebibliography}{} 
%
\bibitem{} Bolatto A.D., Jackson J.M., Israel F.P., Zhang X. $\&$ Kim S., 2000
           \apj 545, 234
\bibitem{} Bolatto A.D., Leroy A., Israel F.P., Jackson J.M., 2003 \apj 
           submitted
\bibitem{} Braun J.M., Bomans D.J., Will J.-M. $\&$ de Boer K.S., 1997 
           \aua 328, 167
\bibitem{} Caldwell D.A. $\&$ Kutner M.L. 1996 \apj 472, 611
\bibitem{} Davies R.D., Elliott H.K. $\&$ Meaburn J., 1976 \mnras 81, 89
\bibitem{} Filipovic M.D., White G.L., Haynes R.F., et al.,\, 1996 \auas 
           120, 77
\bibitem{} Filipovic M.D., Bohlsen, T., Reid, W. et al.,\, 2002 \mnras 
           335, 1085
\bibitem{} Garay G., Johansson L.E.B., Nyman L.-$\AA$, et al.\, 2002 
           \aua 289, 977 (Paper VIII)
\bibitem{} Henize H., 1956 \apjs 2, 315
\bibitem{} Heydari-Malayeri M. $\&$ Lecavelier des Etangs A., 1994 \aua 291, 960
\bibitem{} Heydari-Malayeri M., Charmandaris V., Deharveng L., Rosa M.R. 
           $\&$ Zinnecker H., 1999 \aua 347, 841
\bibitem{} Israel F.P., Johansson L.E.B., Lequeux J., et al.\, 1993 
           \aua 276, 25 (Paper I)
\bibitem{} Israel F.P., de Graauw Th., Johansson L.E.B., et al.\, 2003 
           \aua 401, 991 (Paper IX)
\bibitem{} Johansson L. E. B., Olofsson H., Hjalmarson A., Gredel, R. $\&$
           Black, J. H., \aua 291, 89
\bibitem{} Johansson L.E.B., Greve A., Booth R.S., et al.\, 1998 \aua 331, 857
           (Paper VII)
\bibitem{} Kutner M.L., Rubio M., Booth R.S., et al.\, 1997 \auas 122, 255
           (Paper VI)
\bibitem{} Lequeux J., Le Bourlot J.L., Pineau des For\^ets G., et al.\, 1994
           \aua 292, 371
\bibitem{} Lucke P.B. $\&$ Hodge, P.W., 1970 \aj 75, 171
\bibitem{} McKibben Nail V. $\&$ Shapley H., 1953 Proc. Nat. Acad. Sci. 39, 358
\bibitem{} McLaren I., Richardson K.M. $\&$ Wolfendale A.W., 1988 \apj 333, 82
\bibitem{} Meaburn J., 1980 \mnras 192, 365
\bibitem{} Meaburn J., Laspias V., Solomos N. $\&$ Goudis C., 1989 \aua 
           225, 497
\bibitem{} Mizuno N., Yamaguchi R., Mizuno A., et al.\, 2001 \pasj 53, 971
\bibitem{} Olsen K.A.G., Kim S. $\&$ Buss J.F., 2001, \aj 121, 3075
\bibitem{} Rubio M., Lequeux J., Boulanger F., et al.\, 1993a, \aua 271, 1
\bibitem{} Rubio M., Lequeux J., Boulanger F., et al.\, 1996, \auas 118, 263
\bibitem{} Rubio M., Contursi A., Lequeux J. , et al.\, 2000 \aua 359, 1139
\bibitem{} Schwering P.B.W. $\&$ Israel F.P., 1990, Atlas and 
           Catalogue of Infrared Sources in the Magellanic Clouds, 
           Kluwer, Dordrecht
\bibitem{} Yamaguchi R., Mizuno N., Onishi T., Mizuno A. $\&$ Fukui Y.,
           2001a \pasj 53, 959
\bibitem{} Yamaguchi R., Mizuno N., Onishi T., Mizuno A. $\&$ Fukui Y.,
           2001b \apjl 552, L185

\end{thebibliography}
\end{document}